\pdfoutput=1

\documentclass[twocolumn]{aastex631}

\newcommand*\smallsig[2]{%
  \begingroup\raisebox{.7ex}{\color{#2}\sffamily\bfseries\scriptsize#1}\endgroup}
\newcommand*\largesig[2]{%
  \begingroup\color{#2}\sffamily\bfseries#1\endgroup~}

\newcommand*\DeclareAddCommand[2]{%
  \expandafter\newcommand\csname add#1\endcsname[1]
  {\smallsig{#1}{#2!25}\begingroup\color{#2}##1\endgroup}}
\newcommand*\DeclareDeleteCommand[2]{%
  \expandafter\newcommand\csname del#1\endcsname[1]
  {\smallsig{#1}{#2!25}\begingroup\color{#2!25}##1\endgroup}}
\newcommand*\DeclareReplaceCommand[2]{%
  \expandafter\newcommand\csname rep#1\endcsname[2]
  {\smallsig{#1}{#2!25}\begingroup\color{#2!25}##1\color{#2}##2\endgroup}}
\newcommand*\DeclareNoteCommand[2]{%
  \expandafter\newcommand\csname note#1\endcsname[1]
  {\begingroup\color{#2}[\kern-.15em[\largesig{#1}{#2}##1]\kern-.15em]\endgroup}}
\newcommand*\DeclareEditCommands[2]{%
  \DeclareAddCommand{#1}{#2}%
  \DeclareDeleteCommand{#1}{#2}%
  \DeclareReplaceCommand{#1}{#2}%
  \DeclareNoteCommand{#1}{#2}}

\DeclareEditCommands{EC}{green!70!black}
\DeclareEditCommands{TB}{magenta}

\DeclareEditCommands{VT}{red}

\usepackage{CJK}
\usepackage{amsmath}

\shorttitle{RMHD Simulation of sub-Eddington Circumbinary Disk around an MBHB}
\shortauthors{Tiwari et al.}

\begin{document}
\begin{CJK*}{UTF8}{gbsn}

\title{Radiation Magnetohydrodynamic Simulation of sub-Eddington Circumbinary Disk around an Equal-mass Massive Black Hole Binary}

 \correspondingauthor{Vishal Tiwari}
 \email{vtiwari@gatech.edu}

  \author[0000-0002-7110-9885]{Vishal Tiwari}
  \author[0000-0001-5949-6109]{Chi-Ho Chan}
  \author[0000-0002-7835-7814]{Tamara Bogdanovi\'c}
  \affiliation{Center for Relativistic Astrophysics and School of Physics,
  Georgia Institute of Technology, Atlanta, GA 30332, USA}
  \author[0000-0002-2624-3399]{Yan-Fei Jiang (姜燕飞)}
  \affiliation{Center for Computational Astrophysics, Flatiron Institute,
  162 Fifth Avenue, New York, NY 10010, USA}
  \author[0000-0001-7488-4468]{Shane W. Davis}
  \affiliation{Department of Astronomy, University of Virginia,
  Charlottesville, VA 22904, USA}
  \author{Simon Ferrel}
\affiliation{Center for Relativistic Astrophysics and School of Physics,
  Georgia Institute of Technology, Atlanta, GA 30332, USA}

\begin{abstract}
\noindent
We present the first three-dimensional radiation magnetohydrodynamic (RMHD) simulation of a sub-Eddington circumbinary disk (CBD) around an equal-mass massive black hole binary (MBHB) with a total mass of $2\,\times\,10^7\,M_{\odot}$ on a circular orbit, separated by 100$\,GM_{\rm tot}/c^2$. The inclusion of radiation leads to a denser, thinner, and more filamentary disk compared to non-radiative magnetohydrodynamic simulation, primarily due to reduced pressure support and an altered equation of state. The RMHD disk also features $\sim 3$ times lower accretion rate ($\approx 0.15\,\dot{M}_{\rm Edd}$), weaker accretion streams and a less pronounced overdensity (a.k.a., ``lump") at the inner edge.  Our analysis of the light curves and thermal spectra reveals that the variability induced by the binary-CBD interaction is distinguishable in the optical/UV band, where CBD shines at about $1\%$ of the Eddington luminosity. These findings underscore the crucial role of radiation on the structure and observational properties of CBDs around massive black hole binaries and have implications for detecting electromagnetic counterparts to LISA gravitational wave precursors, and for heavier binaries that are Pulsar Timing Array sources.


\end{abstract}


\section{Introduction}
\label{sec:intro}

Most massive galaxies host a central massive black hole \citep[MBH;][]{kormendy2013coevolution}. When galaxies merge, their MBHs move from thousands of parsecs to within 1-10 parsecs through dynamical friction \citep[e.g.,][]{mayer2013massive,li2020pairing}. Some fraction of MBHs then form a bound binary, which tightens through further interactions with stars and gas and possibly other MBHs, and merge due to the emission of gravitational waves \citep[GWs;][]{begelman1980massive}.

Recent detection by the Pulsar Timing Arrays PTAs of a low-frequency GW background consistent with a population of MBHBs \citep{agazie2023nanograv_gwbackground, antoniadis2024second, reardon2023search, xu2023searching} and targeted searches for individual MBHBs \citep[e.g.,][]{agazie2023nanograv,agazie2024nanograv} bring renewed urgency to understand this population of sources. Individual MBHBs will also be one of the main targets for future spaced-based GW missions, like the Laser Interferometer Space Antenna \citep[LISA;][]{amaro2017laser, amaro2023astrophysics}. No gravitationally bound or merging MBHBs have yet been conclusively observed however. It is anticipated that some MBHBs in gas-rich environments may emit electromagnetic (EM) signals that might allow us to identify them and study the properties of their accretion flows and host galaxies \citep[see][for a review]{bogdanovic2022electromagnetic}.

A number of works in the literature have utilized hydrodynamic simulations to investigate the properties of close binaries surrounded by the gaseous circumbinary disks (CBDs) in the evolutionary stages before or during the GW-driven inspiral. The gravitationally bound MBHBs have usually been found to create a low-density cavity in the CBD. The cavity contains the mini-disks fed by streams originating from the inner edge of the CBD. 
Simulations of black hole binaries with similar masses have identified an additional feature: an overdensity, often called a ``lump," that forms at the inner edge of the CBD. The lump forms when flung-out streams shock and deposit material at the inner edge of the CBD, influencing the mass accretion rate and leaving distinct signatures in the EM light curves of comparable-mass MBHBs. These salient features of binary accretion flows have been identified in a wide range of hydrodynamical (HD) simulations \citep{armitage2005eccentricity,macfadyen2008eccentric,cuadra2009massive,roedig2011limiting,roedig2012evolution,d2013accretion,roedig2014migration,farris2014binary,farris2015binary,farris2015characteristic,nixon2015resonances,young2015binary,ragusa2016suppression,d2016transition,munoz2016pulsed,tang2017orbital,thun2017circumbinary,ryan2017minidisks,miranda2017viscous,tang2018late,munoz2019hydrodynamics,hirsh2020cavity,heath2020orbital,ragusa2020evolution,munoz2020circumbinary,tiede2020gas,duffell2020circumbinary,franchini2021circumbinary,d2021orbital,zrake2021equilibrium,sudarshan2022cooling,smallwood2022accretion,westernacher2022multiband,dittmann2022survey,wang2023role,lai2023circumbinary,franchini2023importance,krauth2023disappearing,dittmann2023decoupling, siwek2023preferential,turpin2024orbital,tiede2024eccentric,franchini2024emission} and magnetohydrodynamical (MHD) simulations \citep{shi2012three,noble2012circumbinary,farris2012binary,giacomazzo2012general,gold2014accretion,gold2014accretion_two,shi2015three,kelly2017prompt,bowen2017relativistic,bowen2018quasi,bowen2019quasi,paschalidis2021minidisk,armengol2021circumbinary,cattorini2021fully,noble2021mass,cattorini2022misaligned,combi2022minidisk,avara2023accretion,bright2023minidisc,ruiz2023general,most2024magnetically}. 

Many of the simulations by necessity make assumptions about the gas thermodynamics, using either an isothermal or adiabatic equation of state (EOS) with simple heating and cooling prescriptions.
These simplifications make simulations of CBDs computationally feasible but they also introduce uncertainties in the properties of the disk and the orbital evolution of the binary.
For example, \citet{sudarshan2022cooling} and \citet{wang2023role, wang2023hydrodynamical} used 2D HD $\beta$-disk models of CBDs to study how cooling and heating affect disk structure and mass accretion rate. They found that longer disk cooling times, compared to instantaneous cooling (implied by the isothermal equation of state), lead to a more symmetrical disk and the disappearance of the low-frequency mode in the mass accretion rate, which was originally produced by the inner-edge overdensity in isothermal disks. Furthermore, the thermodynamics of the CBD, determined by the gas heating and cooling, is known to influence binary orbital evolution: warmer/thicker disks cause binaries to outspiral, while cooler/thinner disks drive inspirals \citep[e.g.,][]{tiede2020gas,dittmann2022survey,lai2023circumbinary}. 

Even though thermodynamics of the gas around the MBHB has been found to have important effects on the binary orbital evolution, accretion rate, and potentially EM signatures, there are presently no simulations of CBDs in which the equation of state is modeled from the first principles, by accounting for interactions of matter and radiation in their magnetized gas. This is largely because the computational cost of such simulations has so far been prohibitive. Effects of radiation have however been studied in radiation magnetohydrodynamic (RMHD) simulations of sub- and super-Eddington accretion disks around single MBHs \citep[e.g.,][respectively]{jiang2019global,jiang2019super}. These simulations showed that radiation can impact the structure and angular momentum transport in accretion disks around MBHs, with potentially important implications for MBHB accretion flows and their observational signatures. 

Motivated by these considerations, we present the first ever three-dimensional, global RMHD simulation of a sub-Eddington CBD around an equal mass $2\times10^{7} M_{\odot}$ MBHB on a circular orbit. We investigate how radiation affects the structure of the CBD, its accretion rate and observational signatures. Our simulations extend over about 75 binary orbits, corresponding to about nine orbits of a fluid element located at the radius $400\,r_{\rm g}$ in the CBD. This duration is set primarily by computational expense of the simulation and is sufficient to capture the evolution of the disk's density profile, mass accretion onto the binary, and the resulting EM emission out to $400\,r_{\rm g}$. It is however insufficient to follow the slower processes, such as the precession of the inner cavity or the torques that govern the long-term evolution of the binary orbit.

The remainder of this paper is organized as follows: in section \ref{sec:method}, we describe the simulation setup, in \ref{sec:results} we describe the results of our simulations and evaluate the emission properties of CBD in section \ref{sec:obs_props}. We discuss implications of our simulations in section \ref{sec:discussion} and conclude in \ref{sec:con}.

\section{Methods}
\label{sec:method}

\subsection{Binary Configuration}
\label{subsec:physical_sys}

We consider a system with a total mass of $M_{\rm tot} = 2\times10^7\, M_{\odot}$ because such systems could produce luminous EM emission and evolve into GW sources for LISA \citep{amaro2017laser,amaro2023astrophysics}.

We simulate an MBHB with mass ratio $q=1$, where $q = M_2/M_1 \leq 1$ is the ratio of the secondary to primary MBH and $M_1+M_2 = M_{\rm tot}$. We focus on a circular MBHB system separated by $a = 100\,r_{\rm g}$ (where $r_{\rm g} = GM_{\rm tot}/c^2$ is the gravitational radius), which is comparable to and slightly larger than a separation at which they may enter the LISA frequency band. For a binary mass of $2\times10^7\,M_{\odot}$, this separation corresponds to $a\approx 4.8 \times 10^{-5}$\,pc, orbital period $t_{\rm orb}$ = $2 \pi (GM_{\rm tot}/a^3)^{-1/2}$ = 7.16 days, and a GW frequency of $f_{\rm GW} = 6.5 \times 10^{-6}$ Hz. Because the time to coalescence due to the emission of GWs \citep[calculated following][]{peters1964gravitational} is $t_{\rm{gw}}$ $\approx$ 24.5 years $\gg t_{\rm orb}$, we neglect orbital evolution due to the emission of GWs and assume that binary orbital separation is fixed over the length of our simulations, which span about 70 binary orbits.

Population synthesis studies suggest that MBHBs occupy a broad parameter space, with total masses from $\sim 10^5 - 10^{10} M_{\odot}$, mass ratios of $\sim 10^{-4}-1$, and accretion rates ranging from sub to super-Eddington \citep{Kelley_2017MNRAS.464.3131K,Katz_2020MNRAS.491.2301K}. Therefore, our setup is best viewed as one representative scenario for MBHB that could produce notable EM signatures and eventually appear in the LISA band. Our setup also has implications for higher mass binaries that are relevant for Pulsar Timing Arrays.

The size of the individual mini-disks around the MBHs is limited by the size of their Roche lobes  \citep{eggleton1983approximations}:
\begin{equation}
    \frac{R_{\rm L2}}{a} = \frac{0.49 q^{2/3}}{0.6  q^{2/3} + \text{ln}(1 + q^{1/3})},
\end{equation}
where $R_{\rm L2}$ is the Roche lobe radius of the smaller MBH and $R_{\rm L2}/a = 0.39$ for $q=1$.
Therefore, the individual mini-disks would extend from ISCO, i.e., $3\,r_{\rm g}$ for a non-spinning MBH (note that $r_{\rm g}$ is defined using the total mass of the system in this paper), and have the outer radius smaller than $39\,r_{\rm g}$. At the same time, the inner edge of the CBD resides at about $2 a = 200\,r_{\rm g}$. Since our primary focus is on the CBD, we excise and do not simulate the central region inside $80\,r_{\rm g}$ occupied by the mini-disks and their individual MBHs. This radius therefore marks the inner boundary of the computational domain.

We use a Newtonian gravitational potential for the binary (equation~\ref{eq:binary}):
\begin{equation}
\label{eq:binary}
\begin{array}{l}
\Phi(\mathbf{r}, t)= - \frac{G M_{\text {tot }}}{2}\left(\left\|\mathbf{r}-\frac{a}{2} \hat{\mathbf{e}}(t)\right\|^{-1}+\left\|\mathbf{r}+\frac{a}{2} \hat{\mathbf{e}}(t)\right\|^{-1}\right)

\end{array}
\end{equation}
where in general case $\hat{\mathbf{e}}(t)$ is the unit vector towards the more massive black hole\footnote{For equal-mass binaries, $\hat{\mathbf{e}}$ points to the MBH designated as $M_1$.}, with the coordinate origin is at the origin of the spherical-polar grid. The MBHB is coplanar with the CBD, accreting at a sub-Eddington rate. The mass of the CBD $\ll$ $M_{\rm tot}$, and therefore we neglect self-gravity of the CBD.

\subsection{Equations of Radiative Magnetohydrodynamics}
\label{subsec:rmhd}
The equations of ideal radiative MHD are as follows:
\begin{equation}
\frac{\partial \rho}{\partial t}+\nabla \cdot(\rho \boldsymbol{v})=0
\end{equation}
\begin{equation}
\frac{\partial(\rho \boldsymbol{v})}{\partial t}+\boldsymbol{\nabla} \cdot\left(\rho \boldsymbol{v} \boldsymbol{v}-\boldsymbol{B} \boldsymbol{B}+\boldsymbol{P}^*\right)=-\boldsymbol{S}_{\boldsymbol{\rm r}}(\boldsymbol{p})
\end{equation}
\begin{equation}
\frac{\partial E}{\partial t}+\nabla \cdot\left[\left(E+P^*\right) \boldsymbol{v}-\boldsymbol{B}(\boldsymbol{B} \cdot \boldsymbol{v})\right]=-S_{\rm r}(E)
\end{equation}
\begin{equation}
\frac{\partial \boldsymbol{B}}{\partial t}-\nabla \times(\boldsymbol{v} \times \boldsymbol{B})=0
\end{equation}
Here, $\rho, \boldsymbol{v}, P^*$ and $\boldsymbol{B}$ are the gas density, velocity, total pressure and the magnetic field, where $\boldsymbol{P}^* = (P_{\rm gas} + B^{2}/2) \boldsymbol{I}$ and $\boldsymbol{I}$ is the unit tensor. The total energy density is given by $E=\frac{P_{\rm gas}}{\gamma - 1}+\frac{1}{2} \rho v^2+\frac{B^2}{2}$, where $\gamma =5/3$ is the adiabatic index of the gas. 

The time-dependent radiative transport equation is given by equation~\ref{eq:rad_trans} with the frequency-integrated source term given in equation~\ref{eq:sourceterm}.
\begin{equation}
\label{eq:rad_trans}
\frac{\partial I}{\partial t}+c \boldsymbol{n} \cdot \nabla I=c S_I,
\end{equation}
\begin{equation}
\label{eq:sourceterm}
\begin{array}{l}
S_I \equiv \Gamma^{-3}\left[\rho\left(\kappa_{\rm s}+\kappa_{\rm a}\right)\left(J_0-I_0\right)\right. \\
\left.\quad+\rho\left(\kappa_{\rm a}+\kappa_{\rm \delta P}\right)\left(\frac{a_{\rm r}c T^4}{4 \pi}-J_0\right)\right. \\
\left.\quad+\rho\kappa_{\rm s}\frac{4(T - T_{\rm rad})}{T_{\rm e}}J_0\right]
\end{array}
\end{equation}
Here $I, c, \boldsymbol{n}$ are the frequency-integrated specific intensity, speed of light, and angular direction. $\Gamma(\boldsymbol{n}, \boldsymbol{v})$ is defined as $\gamma(1-\boldsymbol{n} \cdot \boldsymbol{v} / c)$ where $\boldsymbol{v}$ is the flow velocity and $\gamma$ is the Lorentz factor that is used to transform between the lab and the comoving frame of the fluid element. $I_{0}$ and $J_{0}$ are the frequency integrated specific intensity and mean intensity in the comoving frame of the fluid. $T$, $T_{\rm rad}$, and $a_{\rm r}$ are the gas temperature, radiation temperature, and the radiation constant respectively. $\kappa_{\rm s}, \kappa_{\rm a}$ are the scattering and Rosseland mean absorption opacities respectively. 
$\kappa_{\rm \delta P}$ is the difference between Planck mean opacity and Rossland mean opacity. The approximate energy exchange via Compton scattering is represented by the last term in the equation~\ref{eq:sourceterm} \citep{hirose2009radiation,jiang2019super}, where $T_{\rm e}$ is the effective temperature of the electron rest mass given by $T_{\rm e} = m_{\rm e} c^2/k_{\rm B}$, where $m_{\rm e}$ is the rest mass of the electron and $k_{\rm B}$ the Boltzmann constant.

We use Rosseland and Planck mean opacities from the tables of \citet{zhu2021global}, which provide opacity values as a function of density and temperature and include atomic, molecular, and dust opacities at solar metallicity. For CBD, the most relevant components are the atomic opacities which include contributions from bound-bound, bound-free, free-free, and scattering processes, which \citet{zhu2021global} adopted from the Los Alamos National Lab’s new generation opacity table \citep{colgan2016new}.

The radiation and MHD equations are coupled via the energy and momentum source terms $S_{\rm r}(E)$ and $\boldsymbol{S}_{\boldsymbol{\rm r}}(\boldsymbol{p})$, which are related to the zeroth and the first moment of the source term in equation~\ref{eq:sourceterm} \citep{jiang2021implicit}.
\begin{equation}
S_{\rm r}(E) \equiv c \int S_I d \Omega
\end{equation}
\begin{equation}
\boldsymbol{S}_{\boldsymbol{\rm r}}(\boldsymbol{p}) \equiv \int \boldsymbol{n} S_I d \Omega
\end{equation}
Finally, the radiation energy density, radiation flux, and the radiation pressure tensor in the lab frame are just the zeroth, first and second moments of the specific intensity \citep{rybicki1991radiative}:
\begin{equation}
    E_{\rm r} =\frac{1}{c} \int_{4 \pi} I d \Omega
\end{equation}
\begin{equation}
    \mathbf{F_{\rm r}}=\int_{4 \pi} \mathbf{n} I d \Omega .
\end{equation}
\begin{equation}
    \mathbf{P_{\rm r}}=\frac{1}{c} \int_{4 \pi} \mathbf{n n} I d \Omega .
\end{equation}

\subsection{Numerical Setup}
\label{subsec:num_setup}

We employ the radiative version of the Athena++ \citep{stone2020athena++}, a finite-volume code that simultaneously solves the ideal MHD equation and the time-dependent radiative transport equation for specific intensities \citep{jiang2021implicit}.

We model the CBD in spherical-polar coordinates $(r,\theta,\phi)$ around a time-varying binary potential (equation~\ref{eq:binary}) and cut out the inner region of the domain at $80\,r_{\rm g}$ containing the two MBHs and their mini-disks. We note that the inner boundary is within the ``accretion horizon" -- the radius beyond which no material is returned to the CBD inner edge \citep[located at $\approx 100\,r_{\rm g}$;][]{tiede2022binaries}. This may change in the presence of radiation-driven outflows, particularly around super-Eddington mini-disks. Our choice of the equal-mass MBHB and the sub-Eddington CBD is expected to lead to the sub-Eddington accreting mini-disks, so we do not expect them to launch outflows that would impact the circumbinary disk. The mini-disks are however expected to emit radiation, which could lead to additional heating by irradiation (and potentially increased luminosity) of the inner edge of the CBD. Our simulation does not capture this effect. 

The domain extends from $80\,r_{\rm g}$ to $1600\,r_{\rm g}$ in the radial direction, 0 to $\pi$ in polar, and 0 to $2 \pi$ in the azimuthal direction. We use three levels of static mesh refinement in the region $r \in(80\, r_{\rm g},700\,r_{\rm g}) \text{, } \theta \in (1.47,1.67) \text{, and } \phi \in (0,2\pi)$. The effective (maximum) resolution of the grid is $(r \times \theta \times \phi) = (512 \times 768 \times 512)$. The aspect ratio $\Delta r/r \approx \Delta \theta = \Delta \phi/3 = 4.1 \times 10^{-3}$.

The boundary conditions are periodic in the $\phi$ direction and diode in the $\theta$ and $r$ direction. In the diode boundary conditions, we copy the magnetic and velocity fields of the last active cell in the domain into the ghost cell if the velocity vector is outwards. If the velocity vector points into the simulation domain, we set the radial component of the velocity and magnetic fields to zero while copying other field quantities, along with the pressure and density. These boundary conditions ensure that magnetic fields are not spuriously injected into the domain through the inner and outer boundaries.

\subsection{Initial Conditions}

We initialize the MHD simulation with an axisymmetric torus around the binary potential. The torus is defined by five parameters: maximum density of the torus ($\rho_{\rm m}$), the radius of maximum density ($r_{\rm m}$), polytropic constant ($K$), polytropic index ($\eta$), and shear parameter ($s$). The density $\rho(r,\theta)$ and the pressure of the torus $P_{\rm gas}(r,\theta)$ are related by the polytrope: $P_{\rm gas}(r,\theta) = K \rho(r,\theta)^{\eta}$. The orbital velocity at the density maximum is defined by $\textbf{v}(r_{\rm m},\pi/2) = v_{\rm m}\,\hat{\phi}$. At the density maximum, the pressure is also maximum. Therefore, the pressure gradient vanishes, and at the location of maximum density, we have $r \frac{\partial \langle\Phi\rangle}{\partial r} = v_{\rm m}^2$. The orbital velocity follows the profile $\textbf{v}(r,\theta) = v_{\rm m} (\frac{r \sin\theta}{r_{\rm m}})^{(1-s)} \hat{\mathbf{\phi}}$, where $s$ is the shear parameter that should be between 1.5 and 2 for the torus to be hydrodynamically stable.

Using the hydrostatic balance equation \ref{eq:6}, we solve for the constant of integration C for a given set of parameters ($\rho_{\rm m}, r_{\rm m}$, $K$, $\eta$, $s$). Using the constant, we calculate the value of density, pressure, and velocity at every point of the torus \citep[see also][]{papaloizou1984dynamical, hawley2000global}.

\begin{equation}
\label{eq:6}
\text { C }=\langle\Phi\rangle-\frac{|\textbf{v}|^2}{2(1-s)}+\left\{\begin{array}{ll}
K \frac{\eta}{\eta-1} \rho^{\eta-1}, & \eta \neq 1, \\
K \ln \rho, & \eta=1,
\end{array}\right.
\end{equation}
Here, $\langle \Phi \rangle$ is the time-averaged gravitational potential of the binary as shown in Equation \ref{eq:5}, expanded up to the quadrupole term. This expanded potential facilitates a smaller initial transient instability in the torus compared to the scenario with only the monopole term.

\begin{equation}
\label{eq:5}
\langle\Phi\rangle(r,\theta) = -\frac{GM_{\rm tot}}{r} + \frac{GM_{\rm tot}a^2 q (1+3\cos2\theta)}{8(1+q)^2r^3} 
\end{equation}

\begin{figure*}[ht!]
    \centering
    \includegraphics[width=0.8\textwidth]{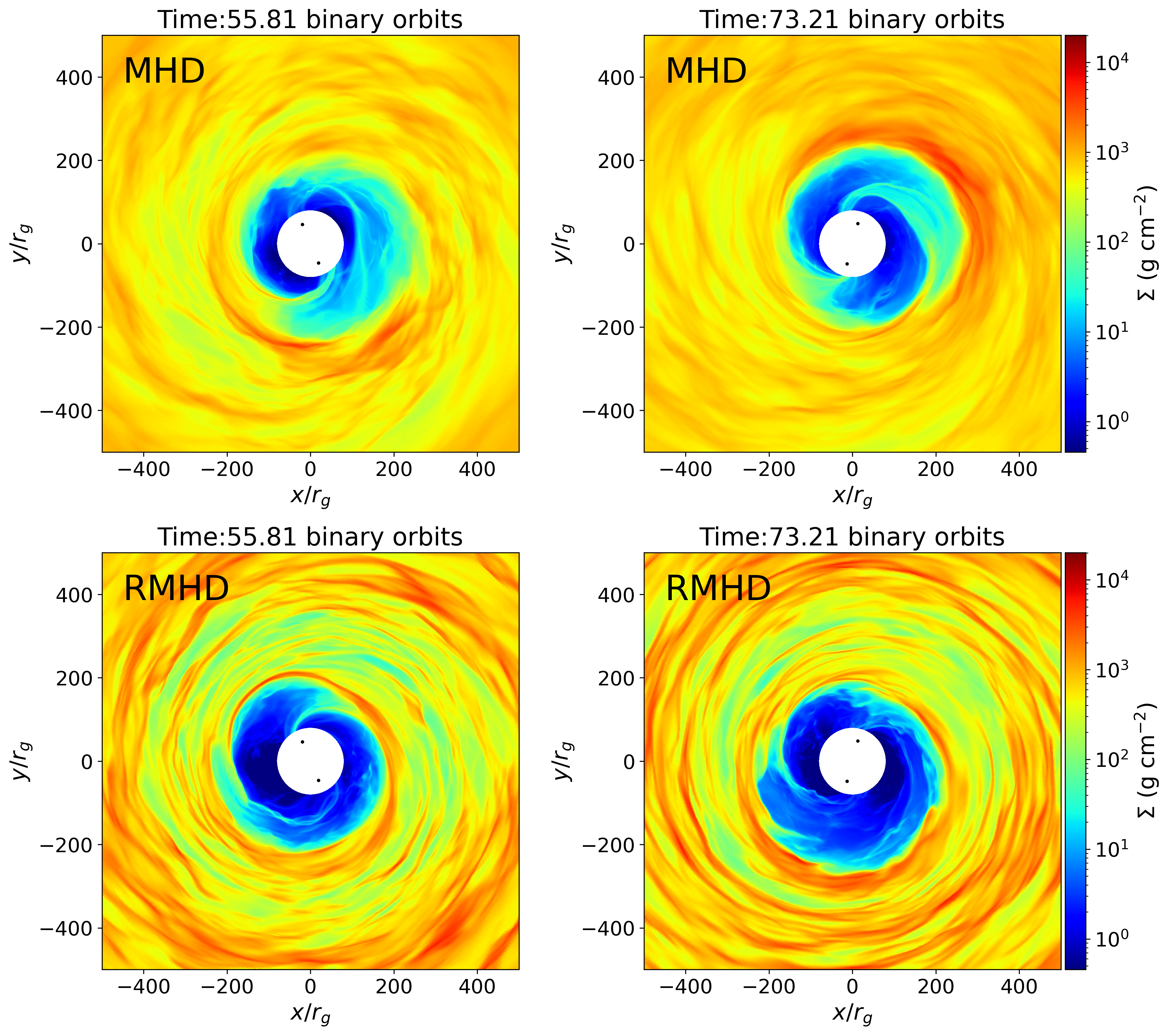}
    \caption{Surface density after $\approx 56$ and $\approx 73$ binary orbits for MHD (top) and RMHD  simulations (bottom). Both show a low-density cavity around the MBHB, approximately $200\,r_{\rm g}$ in size, formed due to binary torques. The RMHD cavity is larger and more depleted of gas. In MHD disk, a distinct lump forms at the CBD's inner edge, whereas in RMHD disk, a thinner ring is present without a prominent lump. RMHD disk also features thin filamentary structures, formed by magnetic pressure compressing the gas. }
    \label{fig:surface_density}
\end{figure*}

We initialize an isothermal torus that leads to the formation of a sub-Eddington CBD with parameters $\rho_{\rm m}=6.17\times10^{-12}$\,g cm$^{-3}$, $r_{\rm m} = 600\,r_{\rm g}$, $\eta = 1$ and $K= 1.09 \times 10^{8}\,$Kelvin and $s = 1.6$. We create a numerical vacuum around the torus with the same prescription as described above but with a different set of parameters with the maximum numerical vacuum density $\rho_{\rm m,\text{vac}} = 3.08 \times 10^{-18}$ g cm$^{-3}$, $r_{\rm m,\text{vac}} = 500\, r_{\rm g}$, $\eta_{\text{vac}} = 1$, $K_{\text{vac}}= 1.09 \times 10^{10}\,$K, and $s_{\text{vac}} = 1.75$. This numerical vacuum helps avoid numerical instabilities that may develop at the edges of the torus.

The magnetic field is defined using the vector potential $A(r,\theta) \propto \text{max}(\rho - \frac{1}{2}\rho_{\rm m},0)$. This vector potential defines concentric single-loop poloidal magnetic fields that coincide with the $r-\theta$ density contours in the torus. The strength of the fields are set so that the average plasma $\beta$ = $P_{\rm gas}/P_{\rm mag}$ = 100. We perturb the pressure by 1\% to seed the magnetorotational instability \citep[MRI;][]{balbus1998instability}.

\subsection{Simulation Strategy}
\label{subsec:sim_strategy}

In order to isolate effects unique to radiation, we conducted two simulations: one MHD with locally isothermal EOS for the gas and the other as RMHD with adiabatic EOS for the gas with $\gamma = 5/3$.

RMHD simulations are an order of magnitude more computationally expensive than those with only MHD. To save computational resources, we started the RMHD simulation from a snapshot of the MHD run at around $t \approx 20$ binary orbits. This specific time was selected for the RMHD starting point because, in the MHD's first 20 binary orbits, the torus is primarily in a transient phase where its inner edge is pushed inwards to form a transient disk, and none of the CBD's notable features have developed yet. By doing this, we effectively save around 20 binary orbits worth of computational effort in the RMHD simulation.

\begin{figure}[t!]
    \centering
    \includegraphics[width=1.0\linewidth]{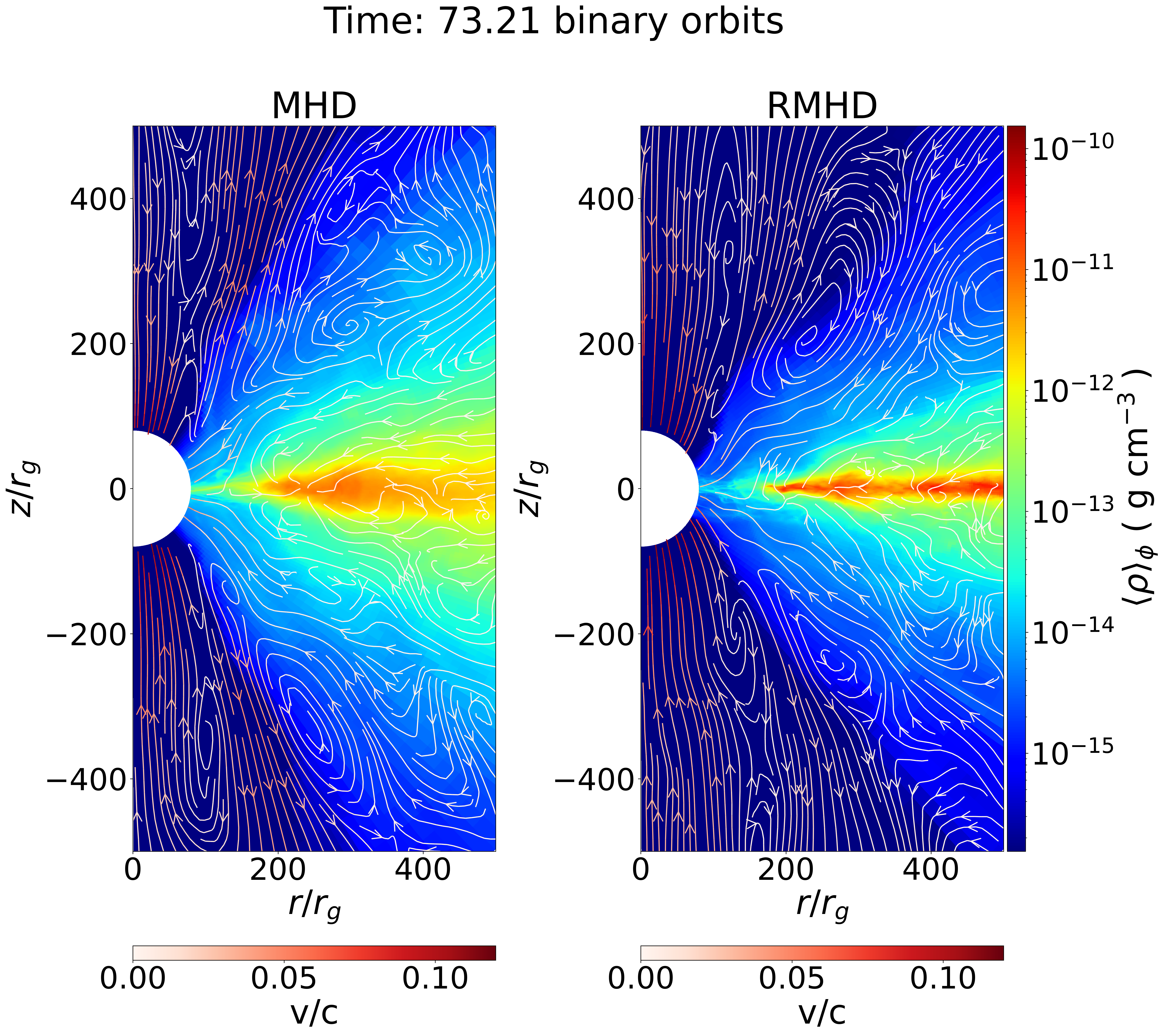}
    \caption{Azimuthally averaged density at t $\approx 73$ binary orbits for the MHD (left) and RMHD (right) simulation. The RMHD disk is a factor of two thinner and denser compared to the MHD disk. The streamlines show the averaged velocity: $\langle V_r \rangle_{\phi}\, \hat{r} + \langle V_{\theta} \rangle_{\phi}\, \hat{\theta}$.
    }
    \label{fig:azimuthal_density}
\end{figure}

In order to incorporate radiation during the restart from the MHD snapshot, we assume that radiation and gas are in local thermodynamic equilibrium (LTE), i.e., that radiation temperature equals the gas temperature ($T_{\rm{rad}} = T_{\rm{gas}}$). Moreover, to maintain the same total pressure in RMHD to the MHD snapshot, we equate the sum of the radiation pressure and gas pressure in RMHD to the MHD gas pressure ($P_{\rm rad} + P_{\rm{gas,RMHD}} = P_{\rm{gas,MHD}}$). These conditions lead to an equation for radiation temperature as shown in equation (\ref{eq:10}).
\begin{equation}
\label{eq:10}
    \frac{a_rT_{\rm{rad}}^4}{3} + \frac{\rho k_{\rm B} T_{\rm{rad}}}{\mu} = P_{\rm{gas, MHD}}
\end{equation}
Here $a_r$ is the radiation constant, $\rho$ is the density of gas, $k_{\rm B}$ is the Boltzmann constant, and $\mu$ is the mean particle mass with $\mu = 0.6\,m_{\rm p}$, where $m_{\rm p}$ is the proton mass. With the gas pressure from the MHD snapshot, we numerically solve for the radiation temperature and set the specific intensities on the angular grid in the lab frame as $I = \frac{\sigma}{\pi} T_{\rm rad}^4$, where $\sigma$ is the Stefan-Boltzmann constant.

We run the MHD simulation for $\approx 75$ binary orbits and the RMHD for $\approx 73$ binary orbits, which corresponds to $\approx$ 9 orbits at a radius of $400\, r_{\rm g}$. We resolve the fastest growing modes of MRI for the MHD and the RMHD runs. A detailed discussion of this matter can be found in Appendix~\ref{subsec:MRI_resolution}.

\section{Results}
\label{sec:results}

Here we summarize the properties of CBDs found in MHD and RMHD simulations and provide more detail in the following subsections. Our simulations produce an RMHD disk that is filamentary, denser, and thinner than the MHD disk (Section~\ref{subsec:disk_structure}). Both disks develop overdensities at the CBD's inner edge but in the RMHD case, the overdensity  (Section~\ref{subsec:Lump}) and the accretion streams (Section~\ref{subsec:streams_cavity}) are less pronounced. MHD stresses (Section~\ref{subsec:stresses}) driven by MRI are weaker in the RMHD disk, causing it to accrete at the rate three times lower than the MHD disk (Section~\ref{subsec:mass_acc_rate}). 

\begin{figure}[t!]
\centering
\includegraphics[width=0.4\textwidth]{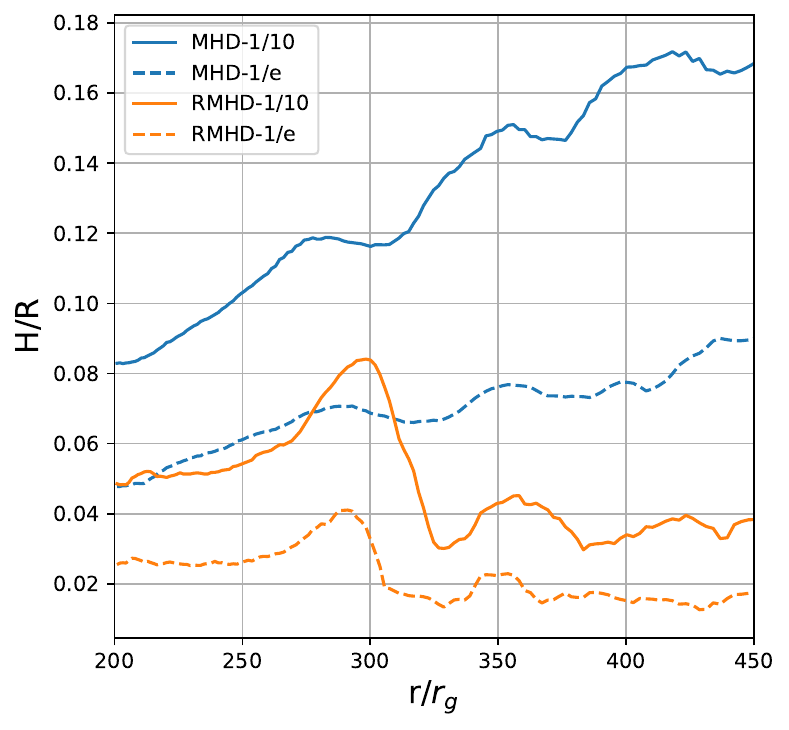}
\caption{\label{fig:scaleHeight} Scale height as a function of radius for the MHD (blue) and the RMHD disk (orange). The curves indicate height where the density above/below the midplane decreases by a factor of 10 (solid curves) or by a factor of $e$ (dashed). The profiles are time averaged over $68-73$ binary orbits for both runs.}
\end{figure}

\subsection{Global Disk Structure and Pressure Support}
\label{subsec:disk_structure}

Figure~\ref{fig:surface_density} shows the surface density of the gas in the circumbinary disk simulated in the MHD and RMHD runs. Binary torques create a low-density cavity of size $\approx 200\,r_{\rm g}$ around the binary, with the RMHD disk cavity being more depleted of gas and slightly larger than in the MHD disk. We find the RMHD disk to be more filamentary than the MHD disk, with the high-density filaments formed in regions of low magnetization.  The presence of a non-axisymmetric overdensity at the inner edge of the CBD is evident in both the MHD and RMHD run but in the latter, the overdensity has a weaker density contrast against the strong density fluctuations of a filamentary background. We discuss the properties of the overdensity in more detail in section \ref{subsec:Lump}.

Figure~\ref{fig:azimuthal_density} shows the corresponding gas density distributions in the plane vertical to the surface of the disk. The most notable difference is that the RMHD disk is visibly thinner and denser than the MHD disk by a factor of approximately two. In both runs, the gas in the bulk of the disk flows inward, as illustrated by the streamlines in Figure~\ref{fig:azimuthal_density}. Above and below the disk the binary (gravitational) torques drive outflows with speeds that are less than the escape velocity, leading to the fallback of gas onto the binary and the disk from the poles.

The vertical structure of the disk is also illustrated in Figure~\ref{fig:scaleHeight}, which shows the scale height profiles of the CBDs calculated from the MHD and RMHD runs. In the MHD run, $H/R$ increases from $\approx\,0.05–0.08$ at $200\,r_{\rm g}$ to $\approx$ 0.09--0.17 at $450\,r_{\rm g}$. In the RMHD disk, the scale height increases up to about $300\,r_g$ (near the lump) before dropping to a nearly constant value in the range $0.02$--$0.04$. The significant difference in the CBD thickness in the MHD and RMHD runs indicate that radiation changes the equation of state of the disk, resulting in a disk with a lower effective sound speed in the RMHD run. As a reference, the \citet{shakura1973black} alpha-disk model predicts $H/R \approx 2\times 10^{-3}$ at 450\,$r_g$ for a $2 \times 10^{7} M_{\odot}$ mass black hole accreting at 15\% of the Eddington rate, showing that the MHD and RMHD simulated disks are considerably thicker than alpha disks, consistent with simulations by \cite{jiang2019global} of accretion disks around single MBHs.

\begin{figure}[t!]
  \includegraphics[trim= 0 0 0 0, clip, width=0.99\linewidth]{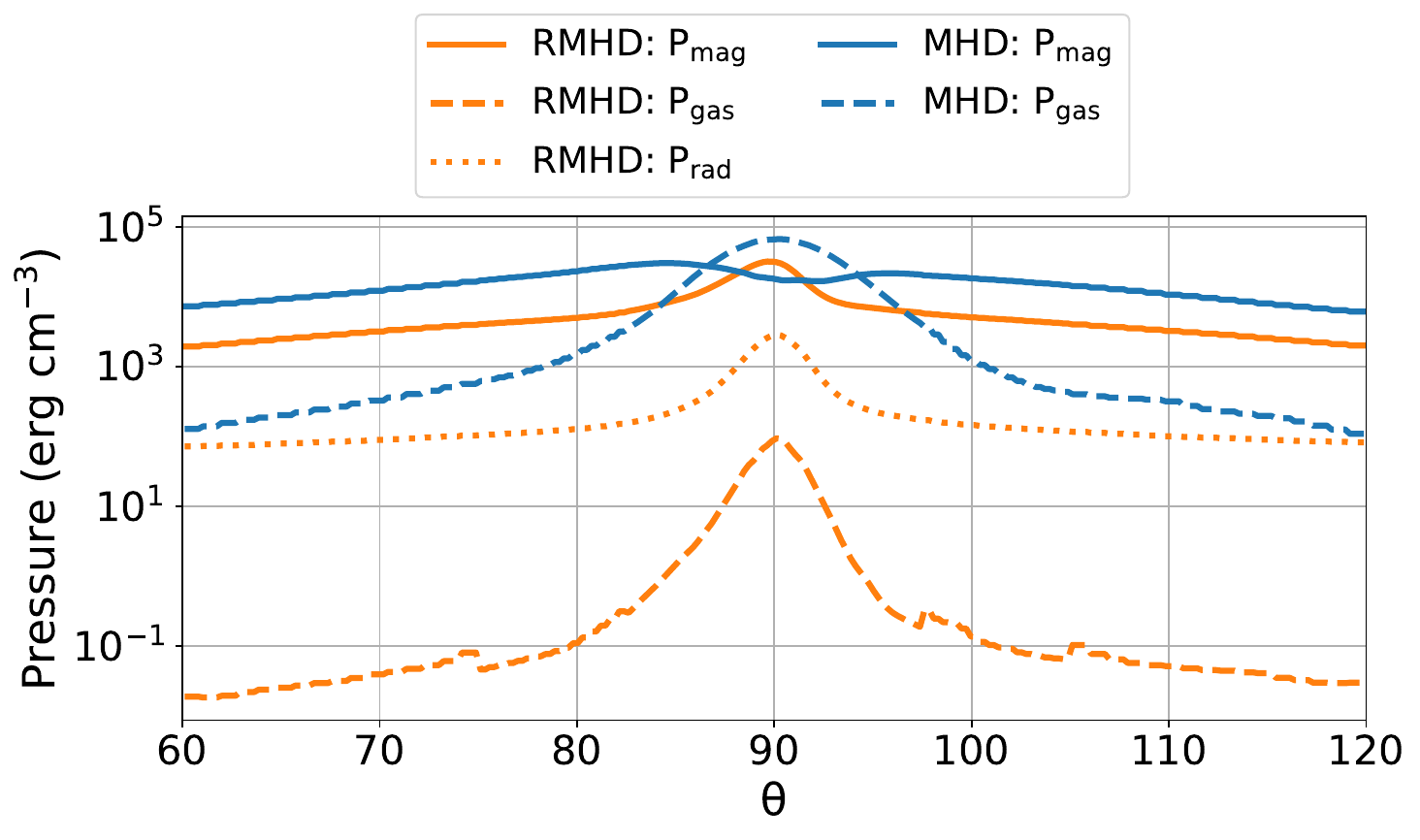}
\caption{
Time and azimuthally averaged magnetic, thermal, and radiation pressure for the RMHD run (orange) and the magnetic  and thermal pressure for the MHD run (blue) for the time window of $68-73$ binary orbits. See the legend for explanation of different line styles.
}
\label{fig:vertical_density_pressure}
\end{figure}

The vertical structure of the CBD discussed in previous paragraphs is determined by the vertical pressure gradient, which supports it against gravity in the vertical direction. In the simulated CBDs this vertical support is provided by a combination of the magnetic, thermal and radiation pressure. This is shown in Figure~\ref{fig:vertical_density_pressure}, which illustrates that at $r=250\,r_{\rm g}$ the MHD disk is predominantly supported by thermal pressure in the midplane of the disk and by magnetic pressure outside of it. The magnetic pressure in the MHD disk dips at the midplane, indicating that magnetic field compresses and confines the gas in vertical direction. This is in contrast with the RMHD disk, which is predominantly supported by magnetic pressure, both in and outside of the disk midplane.

\begin{figure}[t!]
    \centering
    \includegraphics[width=0.48\textwidth]{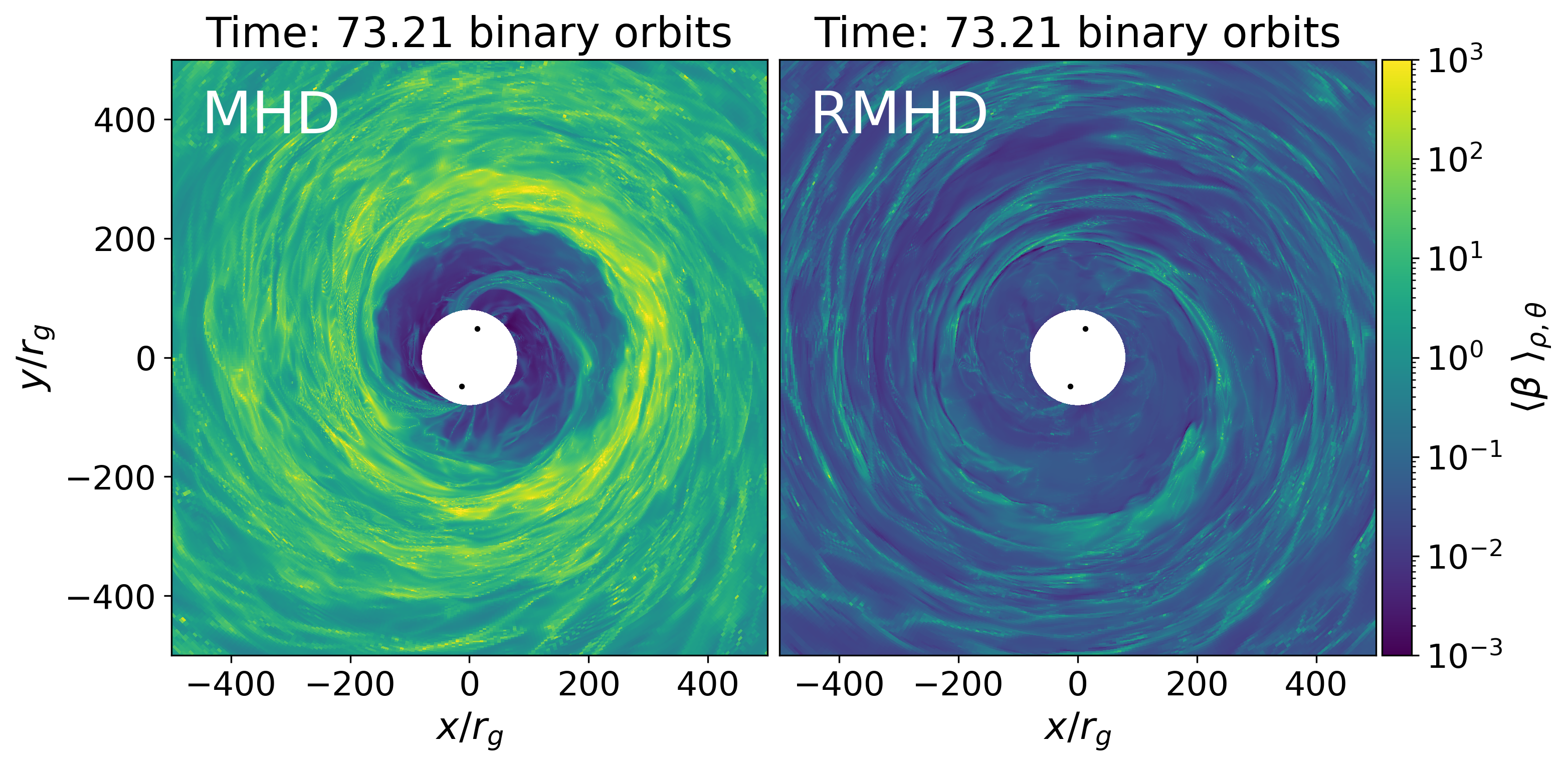}
    \caption{Density-weighted plasma $\beta$ averaged over the polar angle $\theta$ for the MHD and the RMHD runs at $t \approx 73$ binary orbits. The MHD (RMHD) disk is dominated by thermal (magnetic) pressure, and plasma $\beta > 1$ ($\beta<1$).}
    \label{fig:beta}
\end{figure}

These differences between the MHD and RMHD simulated disks is interesting in light of the fact that they started with the same total pressure and the requirement that initially $P_{\rm rad} + P_{\rm{gas, RMHD}} = P_{\rm{gas, MHD}}$ (see Section~\ref{subsec:sim_strategy}). 
Figure~\ref{fig:vertical_density_pressure} however shows that by the end of the simulations the combined radiation plus gas pressure in the RMHD disk is about an order of magnitude smaller than the gas pressure in the MHD disk (for which $P_{\rm rad}=0$). 
This is because radiation escapes from the RMHD disk causing it to lose some of its pressure support. The same description applies to CBD radii beyond $200\, r_g$, as illustrated by the plasma parameter $\beta = (P_{\rm gas}+P_{\rm rad})/P_{\rm mag}$ shown in Figure~\ref{fig:beta}. This indicates that the disk in the RMHD simulation settles into a distinct thermodynamic state relative to its MHD counterpart. Figure~\ref{fig:beta} also shows that besides the global structure of the CBD, magnetic field also shapes its local properties by confining the gas into filaments. The striking filamentary appearance of the RMHD disk is a direct consequence of the local fluctuations in magnetic field strength.

Finally, the dominance of magnetic pressure in our RMHD disk simulation aligns with earlier findings \citep{jiang2019global,skadowski2016three} and corroborates the prediction by \citet{begelman2007accretion} that magnetic support stabilizes disks otherwise prone to thermal instability when radiation pressure dominates over thermal pressure \citep{shakura1976theory}.

\subsection{Overdensity at the Inner Edge of the Disk}
\label{subsec:Lump}

\begin{figure*}[ht!]
    \centering
    \includegraphics[width=1.0\textwidth]{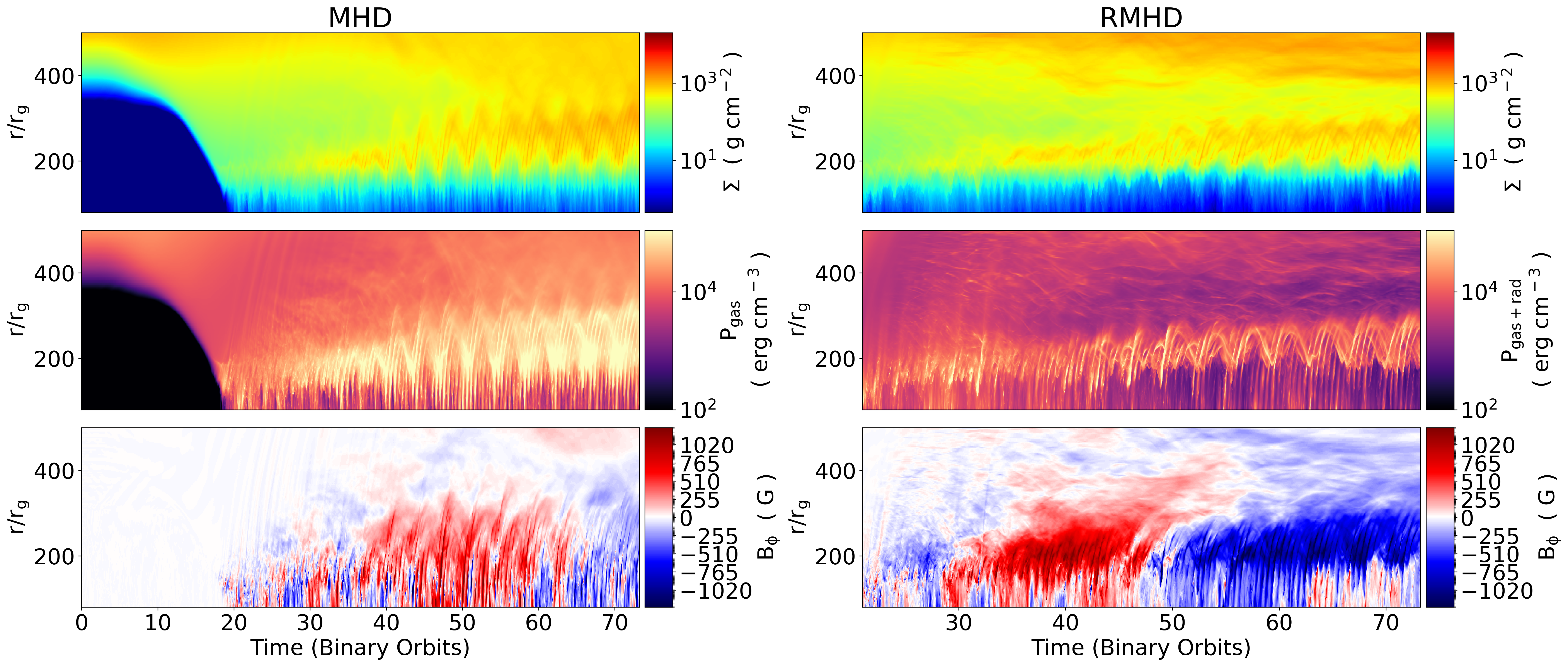}
    \caption{{\it Top:} Radial surface density evolution in the MHD (left) and RMHD run (right). {\it Middle:} Density-weighted gas (or gas + radiation) pressure averaged on shells of constant radii for the MHD (RMHD) disk. {\it Bottom:} Density-weighted $B_{\phi}$ averaged on shells of constant radii for the MHD and RMHD disks.}\label{fig:space_time_Radial_MHD_vs_RMHD}
\end{figure*}

\begin{figure*}[ht!]
    \centering
    \includegraphics[width=1.0\textwidth]{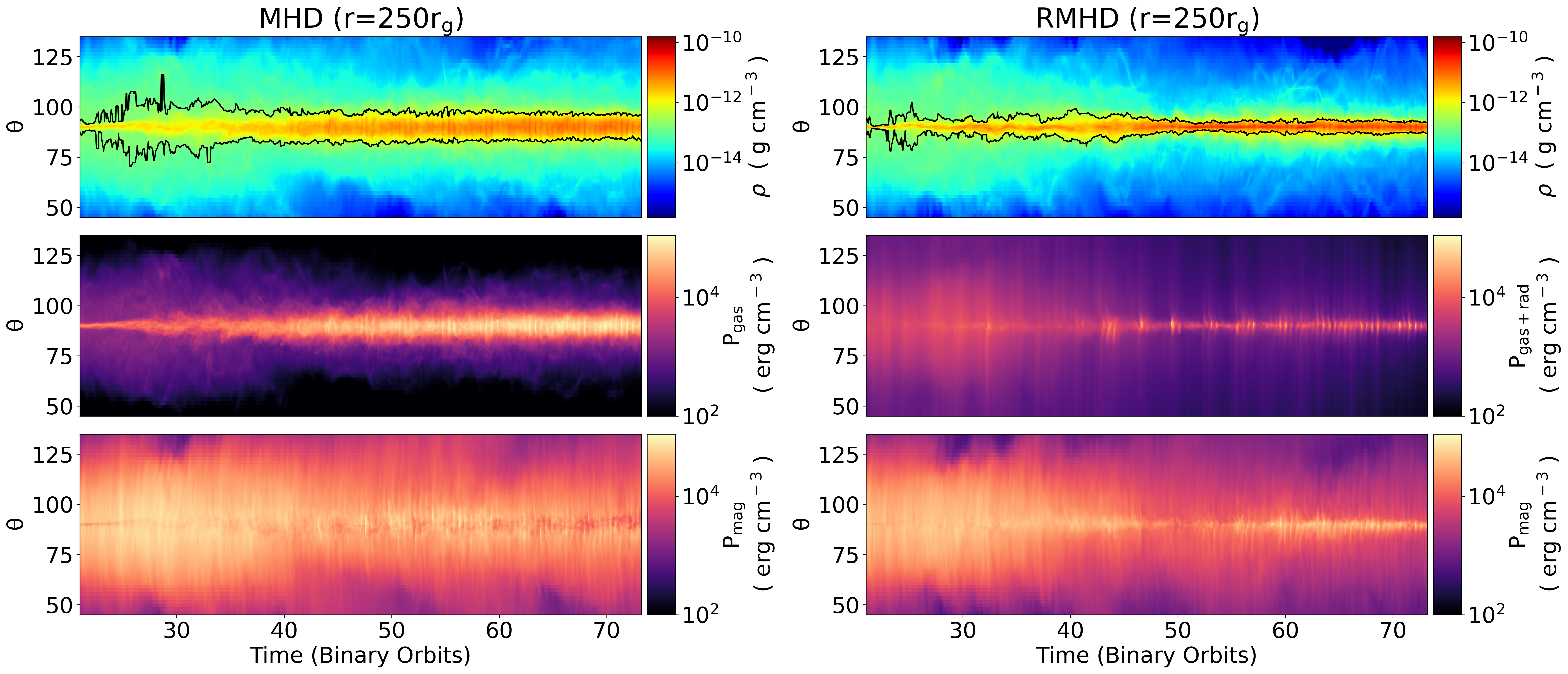}
    \caption{
    Space-time diagrams of the azimuthally averaged density and pressure at a radius of $250\,r_{\rm g}$ (the location of the overdensity) in the MHD (left) and RMHD (right) runs. {\it Top:} Gas density. The RMHD disk is noticeably thinner and denser than the MHD disk, with the black solid line indicating 0.1 times the density of the mid-plane. {\it Middle:} The thermal pressure within the MHD disk exceeds the combined thermal and radiation pressure in the RMHD disk. {\it Bottom:} Magnetic pressure. The MHD run's lump shows weaker magnetization than adjacent regions, whereas the RMHD overdensity exhibits notably stronger mid-plane magnetization compared to above and below the disk.
}\label{fig:space_time_theta_r250_MHD_vs_RMHD}
\end{figure*}

As noted before, both MHD and RMHD simulations show evidence of an overdensity that forms at the inner edge of the CBD (Figure~\ref{fig:surface_density}) but the two appear qualitatively and quantitatively different in different runs.
The MHD overdensity, often referred to as a ``lump,'' is non-axisymmetric, while the RMHD overdensity resembles a thin ring - narrower in radius and more azimuthally extended compared to the MHD lump. The ring-like overdensity in RMHD is also less pronounced against the denser disk background than the distinct lump observed in the MHD run.

The evolution of the overdensity is best understood in the context of the disk. Figure~\ref{fig:space_time_Radial_MHD_vs_RMHD} shows the temporal evolution of the radial profiles of the MHD ($0-73$ binary orbits) and RMHD ($20 - 73$ binary orbits) circumbinary disks. Both runs exhibit an overdensity moving on an eccentric orbit at the CBD's inner edge (visible as the zig-zag pattern) starting from about 30 binary orbits. Throughout most of the simulation, the overdensity and the rest of the disk in the MHD run exhibit noticeably higher thermal pressure and weaker magnetic fields relative to the RMHD run.

\begin{figure*}[t!]
    \centering
    \includegraphics[width=1.0\textwidth]{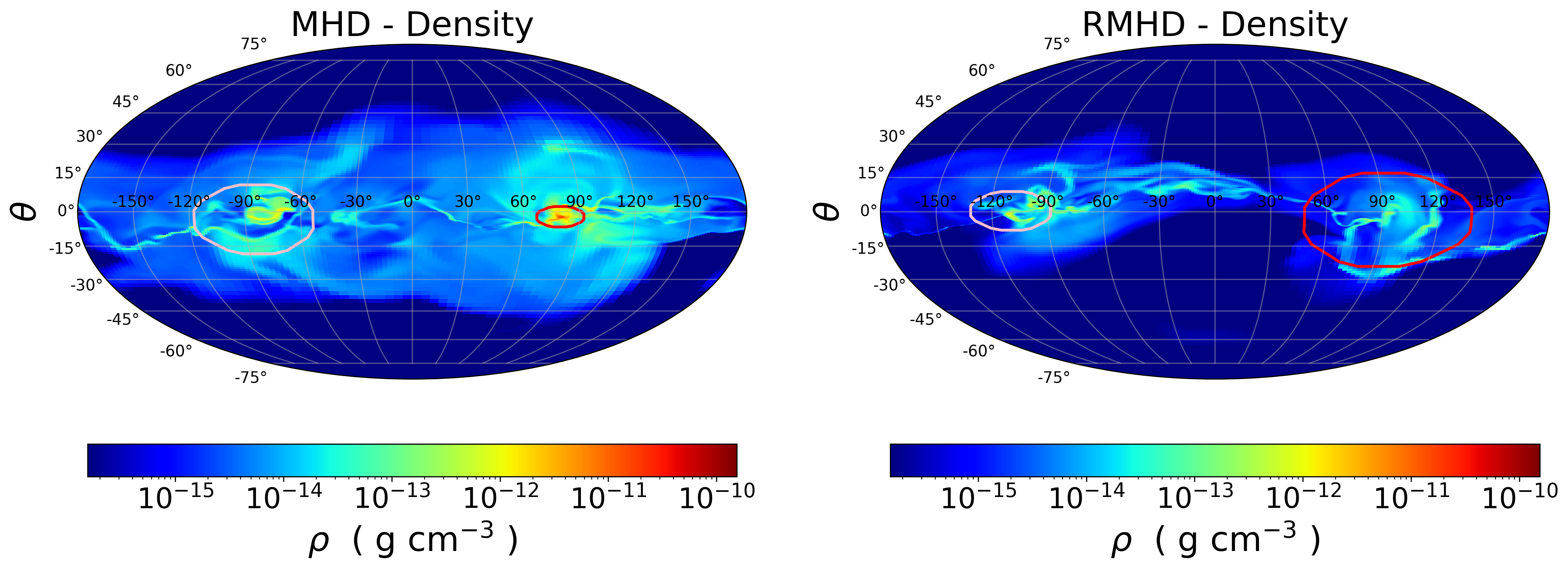}
    \caption{Spherical slices of density at $r=80\,r_{\rm g}$ for the MHD (left) and RMHD (right) simulations at approximately 73 binary orbits. Streams identified using the method described in Section~\ref{subsec:streams_cavity} are marked as pink (falling toward $M_1$) and red (falling toward $M_2$) ellipses. The streams are irregular and are weaker in the RMHD run.}
    \label{fig:streams_props}
\end{figure*}

Figure~\ref{fig:space_time_theta_r250_MHD_vs_RMHD} shows evolution of the density, thermal plus radiation pressure, and magnetic pressure at the location of the overdensity ($r=250\,r_{\rm g}$). The top panels of this figure show that in both runs, the overdensity builds up by about 40 binary orbits, as indicated by the increased density in the midplane of the disk beyond this point in time. The vertical scale height and the density contrast of the MHD lump remain approximately the same in the MHD disk beyond this point. They nevertheless continue to evolve in the RMHD disk between $40-50$ orbits, in which the overdensity region becomes not only denser but also visibly geometrically thinner. This difference is caused by the gradual escape of radiation from the RMHD disk, illustrated by the weakening of the pressure in the middle panel of Figure~\ref{fig:space_time_theta_r250_MHD_vs_RMHD}). The bottom panels show the differences in evolution of the magnetic pressure at the location of the overdensity in the two runs. In the MHD run, magnetic pressure dips in the midplane of the disk (visible as a dark strip that develops after 40 binary orbits), creating a magnetic pressure gradient that confines the overdensity from top and bottom. In contrast, the RMHD disk shows peak magnetic pressure at the midplane, providing support against gravitational collapse of the overdensity.

\subsection{Streams and the Cavity}
\label{subsec:streams_cavity}
\label{subsec:disk_ecc}

Streams are launched from the inner edge of the CBD and flow towards the individual MBHs via the L2 and L3 Lagrange points of the binary potential. We find that the streams that form in both MHD and RMHD run are filamentary with irregular density distributions, as evident in Figure~\ref{fig:streams_props}. To better identify their location, we consider the material as a part of the streams if it falls within an elliptical region centered on each of the two highest points of $\rho v_r$. The major axis of the ellipse in the $\theta$-direction is the full-width at 1/10th maximum of the distribution $\rho v_r (\theta)$ computed using equation \ref{eq:stream_distribution}, and likewise for the $\phi$ direction.

\begin{equation}
\label{eq:stream_distribution}
    \rho v_r (\phi) = \frac{\int \rho v_r d\theta}{\int d\theta} ,\ \ \ \ \ \rho v_r(\theta) = \frac{\int \rho v_r d\phi}{\int d\phi} \;.
\end{equation}

\begin{figure}[t]
\centering
  \includegraphics[trim= 0 0 0 0, clip, width=0.80\linewidth]{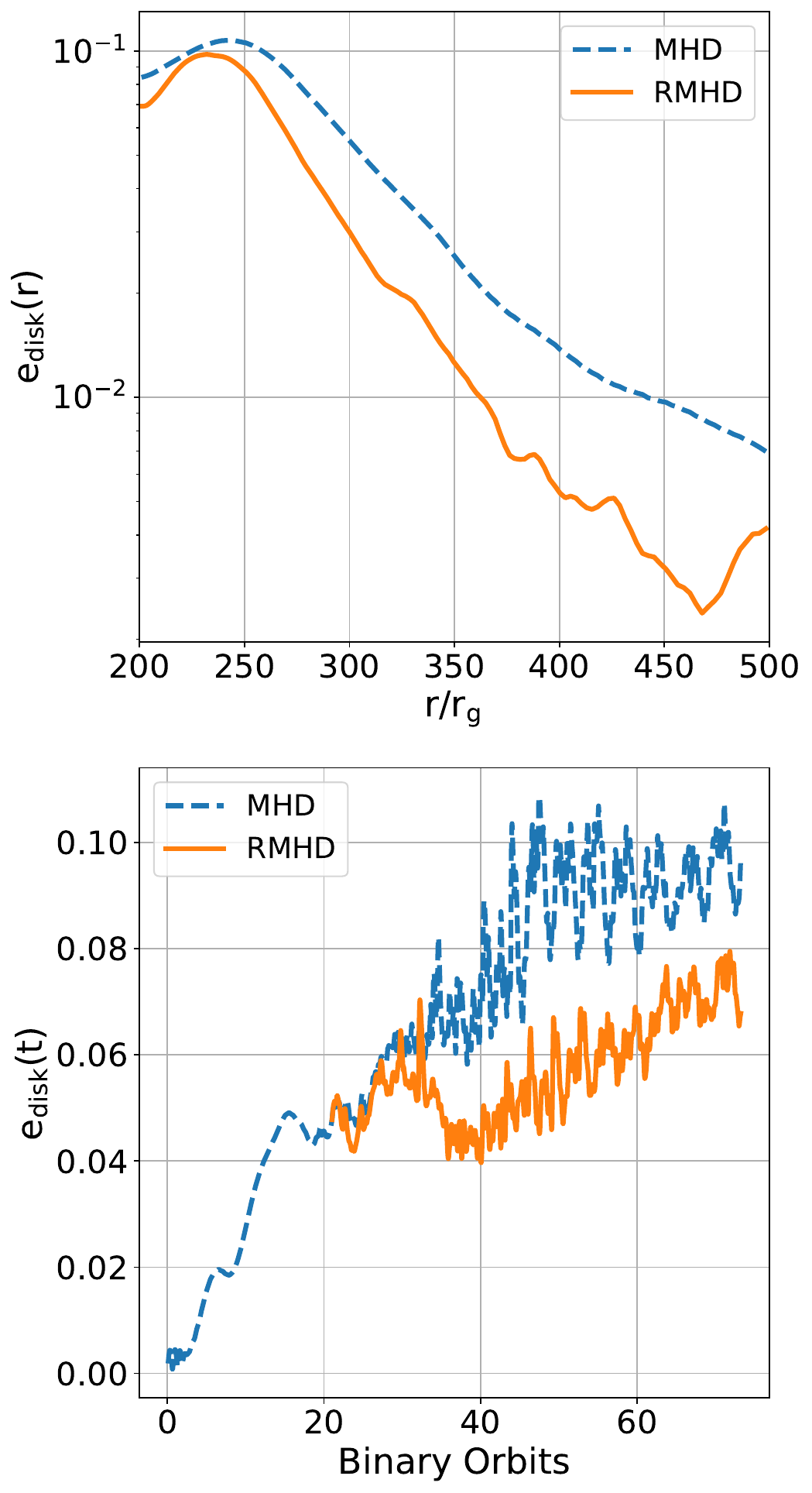}
\caption{{\it Top:} Radial profiles of the disk eccentricity for the MHD (blue) and RMHD (orange) runs averaged over $\sim 68-73$ binary orbits. {\it Bottom:} Time evolution of the eccentricity at the inner edge of the disk ($200-400\,r_{\rm g}$).
}
\label{fig:eccentricity}
\end{figure}

Streams flung out by the binary potential impact the inner edge of the CBD, where they dissipate energy due to shocks and magnetic reconnection and excite eccentricity at the inner edge. The causal relationship between the stream impact and the eccentricity of the CBD at the inner edge has been previously demonstrated by \citet{shi2012three}. They increased the size of the inner edge of their computational domain until the flung-out streams were not captured in the simulation, at which point they recorded no increase in the CBD eccentricity. To further understand the effects of radiation on CBD eccentricity, we compare the radial eccentricity profiles for the MHD and RMHD runs in Figure~\ref{fig:eccentricity}. We compute the radial disk eccentricity as
\begin{equation}
\label{eq:ecc}
e_{\rm disk}(r,t) = \frac{| \langle \rho v_r e^{i\phi} \rangle_{\theta,\phi}|}{\langle \rho v_{\phi} \rangle_{\theta,\phi}} \;.
\end{equation}

Equation~\ref{eq:ecc} evaluates the eccentricity of the CBD as the mean deviation from a perfect circular motion of a fluid element by taking the ratio of the local radial velocity perturbation with the dominant azimuthal velocity. Using this approach, we find that the MHD and RMHD runs have similar eccentricity values of around 0.1 at $r\approx 200\,r_{\rm g}$ and that they diverge at larger radii.


The bottom panel of Figure \ref{fig:eccentricity} shows the disk eccentricity evolution for the inner regions of the CBD, in the range $r = 200-400\,r_{\rm g}$ calculated as 
\begin{equation}
e_{\rm disk}(t) = \frac{|\int_{200\,r_{\rm g}}^{400\,r_{\rm g}} dr \ \langle \rho v_r e^{i\phi} \rangle_{\theta,\phi}|}{\int_{200\,r_{\rm g}}^{400\,r_{\rm g}} dr \ \langle \rho v_{\phi} \rangle_{\theta,\phi}} \;.   
\end{equation}
After 60 binary orbits, the MHD disk eccentricity oscillates between $0.08-0.1$, while the RMHD disk eccentricity keeps growing. The lower eccentricity in the RMHD disk can be attributed to the weaker streams that deliver less momentum and energy to the inner edge of the CBD.

\subsection{RMHD Stresses}
\label{subsec:stresses}

\begin{figure}[t!]
  \centering
  \includegraphics[trim=0 0 0 0, clip, width=0.80\linewidth]{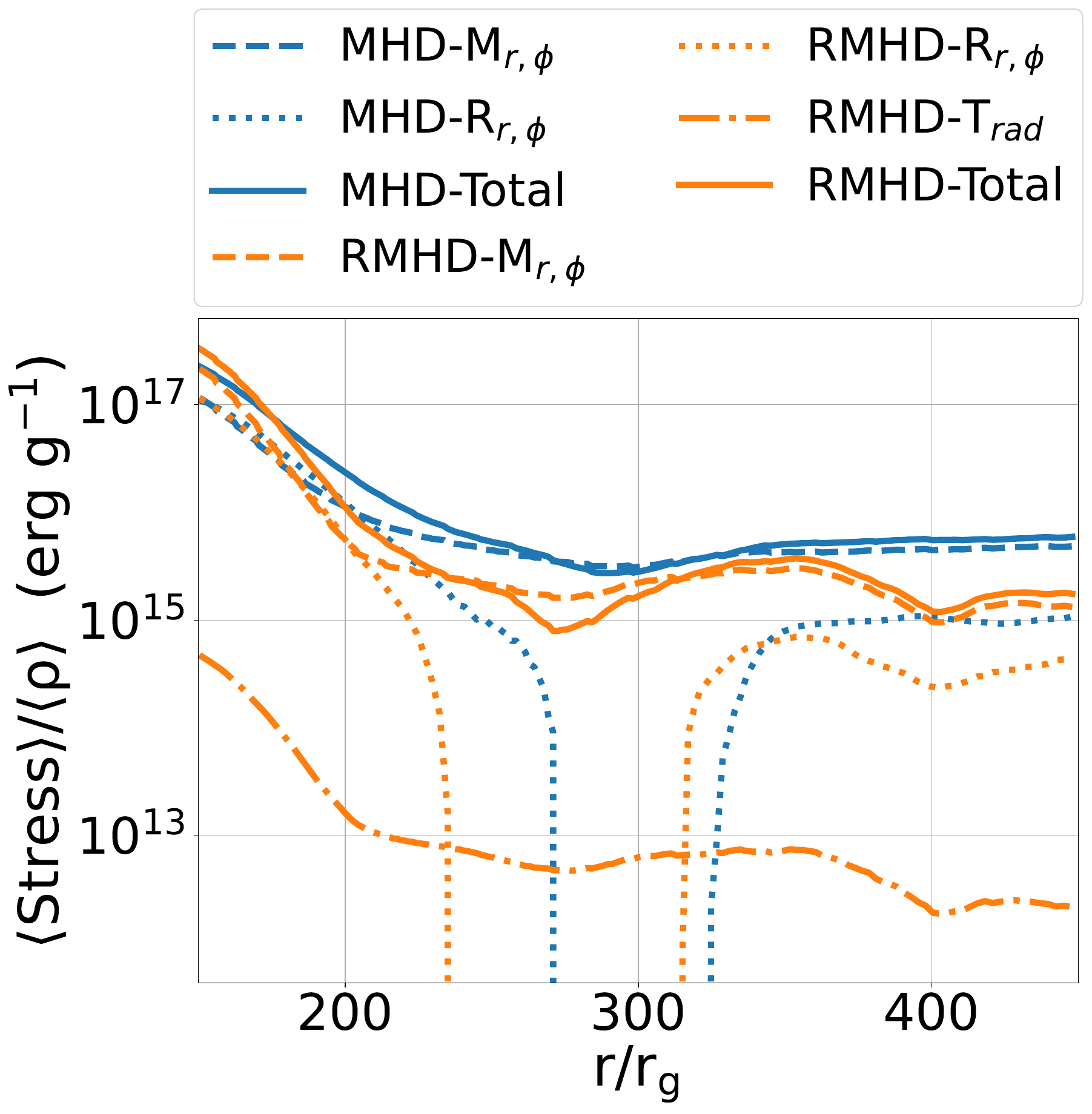}
  \includegraphics[trim= 0 -10 0 0, clip, width=0.8\linewidth]{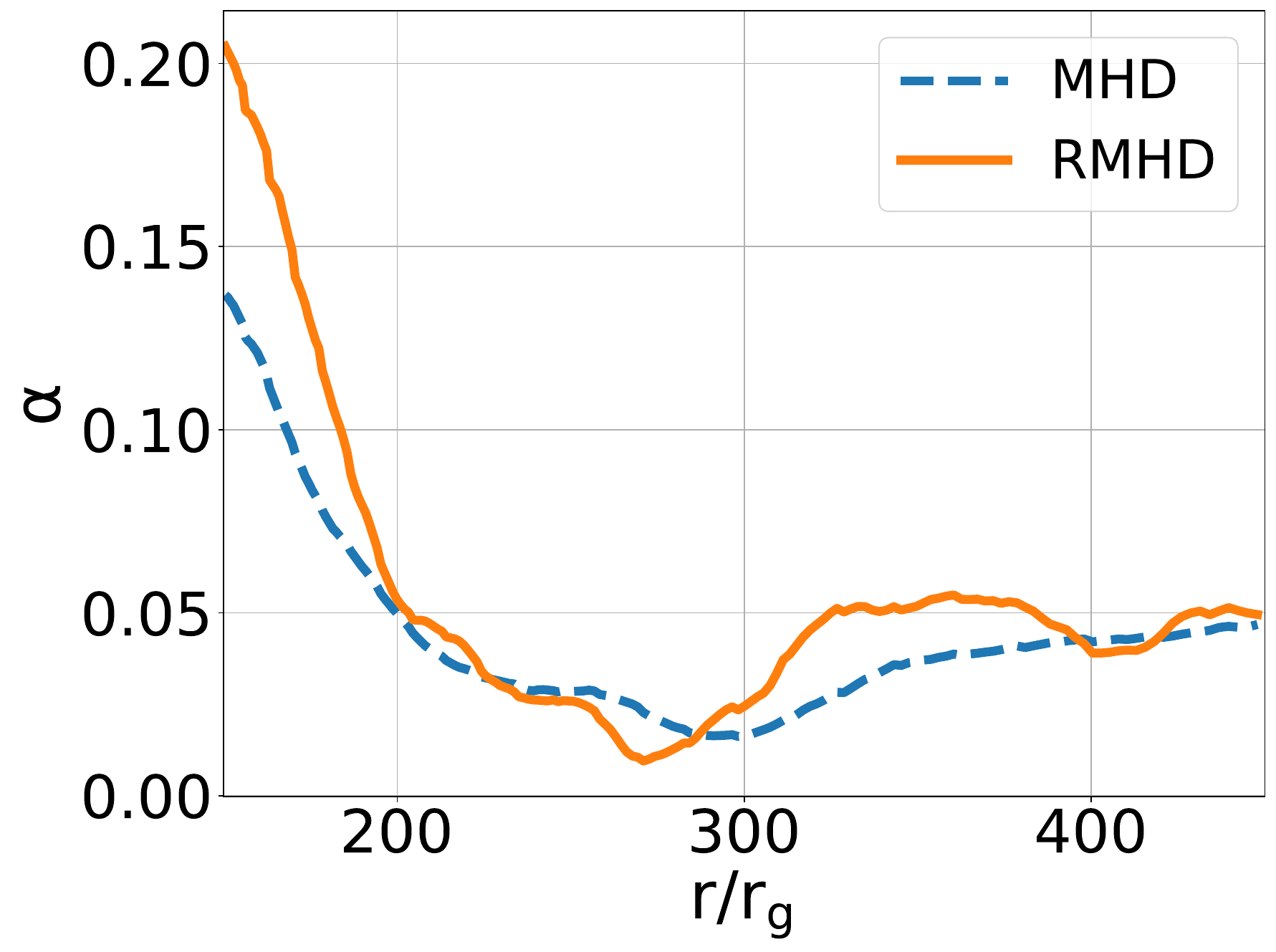}
\caption{{\it Top:} Shell-integrated radial profiles of Maxwell, Reynolds, and radiation stresses in both MHD (blue) and RMHD (orange) simulations. The weak radiation stress play a negligible role in angular momentum transport within the RMHD disk.
{\it Bottom:} Radial profile of $\alpha$ in the MHD and RMHD simulations. Despite the lower total stress in the RMHD disk, the alpha values in RMHD and MHD simulations are comparable. 
}
\label{fig:stress_alpha}
\end{figure}

 To understand what processes drive accretion in the CBD we compare the Maxwell \citep[due to MRI;][]{balbus1998instability}, Reynolds (linked to coherent velocity fluctuations), and radiation stresses in our simulations.

The Maxwell stress, produced by magnetic fields, is computed using equation (\ref{eq:maxwellstress_eq}), where $B_r$ and $B_{\phi}$ are the radial and the azimuthal component of the magnetic field respectively.
\begin{equation}
\label{eq:maxwellstress_eq}
M_{r,\phi} = - \frac{B_r B_{\phi}}{4 \pi} \;\;.
\end{equation}
The Reynolds stress is computed as
\begin{equation}
\label{eq:reynoldsstress_eq}
    R_{r,\phi} = \rho\, \delta v_r\, \delta v_{\phi} \;,
\end{equation}
where $\rho$ is the density and $\delta v_r$ and $\delta v_{\phi}$ are the perturbed velocities $  \delta v_r = v_r - \langle v_r \rangle_{\theta,\phi,\rho}$ and  $\delta v_{\phi} = v_{\phi} - \langle v_{\phi} \rangle_{\theta,\phi,\rho}$ \citep{shi2012three}.
%
%
We also examine the contribution of the radiation stress in the RMHD simulation \citep{huang2023global} 
\begin{equation}
T_{rad} = P_{r,\phi} \sin(\theta) + P_{\theta,\phi} \cos(\theta) \;,
\end{equation}
where $P_{r,\phi}$ and $P_{\theta,\phi}$ are the respective $r,\phi$ and $\theta,\phi$ components of the radiation stress tensor, calculated using the second moment of the specific intensity. 

Figure~\ref{fig:stress_alpha} shows radial profiles for the calculated stresses. The Maxwell stress is $\sim 2-3$ times the Reynolds stress in the main body of the disk for both the MHD and the RMHD simulations. At the inner edge of the CBD, Reynolds stress become more significant and comparable to the Maxwell stress. This is consistent with results of the MHD run presented by \citet{shi2012three}. We also find that the Maxwell stress in the RMHD disk is a few times lower than in the MHD run, indicating slower angular momentum transport in the former. At the same time, the radiation stress is approximately three orders of magnitude smaller than both the Maxwell and Reynolds stresses, resulting in a negligible contribution to the outward transport of angular momentum, in agreement with \citet{jiang2019global}. The Reynolds stress decreases significantly in the region of $\approx 250-320\,r_{\rm g}$, where one part of each stream is flung outward into the circumbinary disk and generates negative stress, while the rest falls back toward the binary resulting in a positive stress. This partial cancellation reduces the Reynolds stress in these regions, consistent with the results from \citet{shi2012three}.

Using the computed stresses we also calculate the effective $\alpha$ viscosity parameter for the CBD, shown it in the bottom panel of Figure~\ref{fig:stress_alpha}. Here, $\alpha$ is defined as the ratio of the total stress (Maxwell + Reynolds + Radiation) to the total pressure (Thermal + Radiation + Magnetic), with zero radiation pressure and radiation stress for the MHD disk. We find the average value of $\alpha$ in the main body of both the MHD and RMHD disks of about $4.5 \times 10^{-2}$, with a noted drop at the inner edge of the CBD at the location of the overdensity ($r= 200-300\,r_{\rm g}$). This decrease signifies a reduction in stresses and a simultaneous rise in total pressure within the overdensity region of both disks. The $\alpha$ values below $200\,r_{\rm g}$ are notably higher than those in the rest of the disk; however, they are not significant, as gravitational torques from the binary predominantly drive accretion in the inner disk regions.

\subsection{Mass Accretion into the Circumbinary Cavity}
\label{subsec:mass_acc_rate}

\begin{figure}
    \centering
    \includegraphics[width=0.47\textwidth]{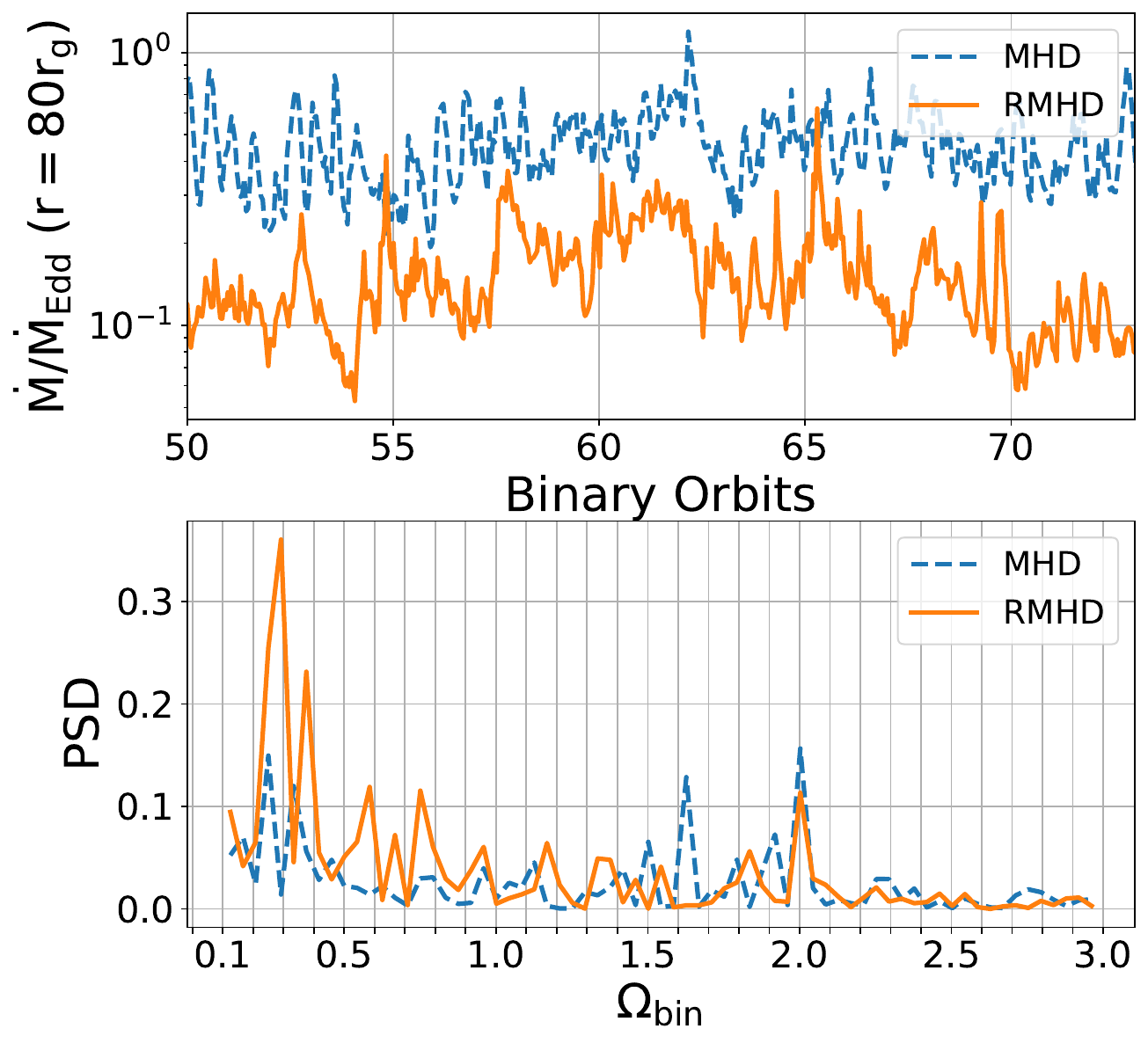}
    \caption{{\it Top:} Total mass accretion rate at the inner edge of the simulation domain for the MHD (blue) and RMHD (orange) runs from approximately 45 binary orbits to the end of the simulations. {\it Bottom:} Power spectral density of the mean normalized mass accretion rate for the same runs. 
    }
    \label{fig:Mdot}
\end{figure}

The infall of matter towards the MBHB in the CBD is driven by the outward transport of angular momentum caused by the stresses developed in the disk. The top panel of Figure~\ref{fig:Mdot} shows the mass accretion rate into the central cutout with $r=80\,r_{\rm g}$, which can be taken as the accretion rate onto both binary members. In the MHD disk, we measure an average mass accretion rate of $0.46\,\dot{M}_{\rm Edd}$, while the RMHD disk averages at $0.15\,\dot{M}_{\rm Edd}$. Here $\dot{M}_{\rm Edd} = L_{\rm Edd}/\eta c^2$, with $L_{\rm Edd}$ being the Eddington luminosity of a $2\times10^7\,M_{\odot}$ source with $\eta = 0.1$. The accretion rate in the RMHD run is more variable, with standard deviation divided by the mean accretion rate resulting in 0.47, compared to the MHD disk at 0.31.

The power spectral density of the mean normalized mass accretion rate is shown in the bottom panel of Figure~\ref{fig:Mdot}. Both the MHD and RMHD simulations show a high-frequency mode at half the binary period ($2\,\Omega_{\rm bin}$), where $\Omega_{\rm bin}$ represents the binary's orbital frequency. This is the consequence of the inner edge of the CBD becoming eccentric and the MBH drawing gas at the point of closest approach twice every orbit \citep{tiede2022binaries}. We also observe low-frequency modes in the range $0.2-0.4\,\Omega_{\rm bin}$ in both runs.
These modes are linked to the orbital motion of the overdensity at the inner edge of the CBD and are consistent with the mode at about $0.25\,\Omega_{\rm bin}$ identified in the MHD simulations described in the literature \citep{noble2012circumbinary}. One notable difference is that the crescent-like overdensity in our RMHD simulation contributes to a range of frequencies, as opposed to one relatively well-defined frequency. This is likely because the fluid elements comprising the overdensity are distributed over orbits with the corresponding range of orbital frequencies. In comparison, our MHD run shows weaker low-frequency modes in a visibly narrower range of frequencies, consistent with $0.25\,\Omega_{\rm bin}$. Furthermore, there is also a peak at the beat frequency of $1.6\,\Omega_{bin}$ in the MHD run, which is absent in the RMHD run.

\begin{figure}[t!]
    \centering
    \includegraphics[width=0.43\textwidth]{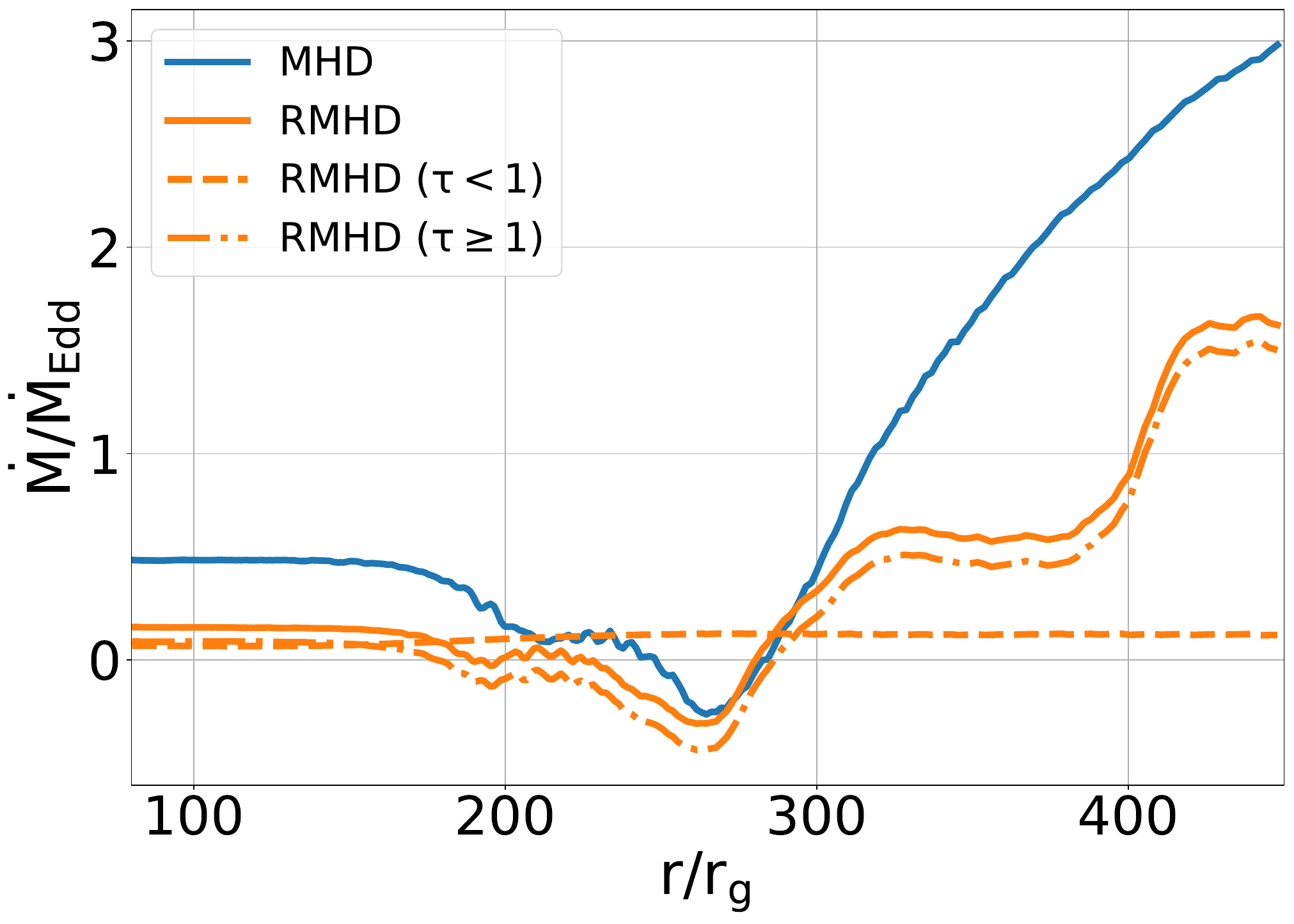}
    \caption{Time-averaged radial profiles of mass accretion rates for the MHD (blue) and RMHD (orange) runs, averaged over $68-73$ binary orbits. The dot-dashed (dashed) line illustrates the contribution to the accretion rate from the optically thick (thin) regions of the RMHD disk.}
    \label{fig:Mdot_Radial_MHD_vs_RMHD}
\end{figure}

Figure~\ref{fig:Mdot_Radial_MHD_vs_RMHD} shows the radial profiles for mass accretion rates measured in the MHD and RMHD runs, calculated as $\dot{M}(r) = \iiint \rho v_r\, r^2 \sin \theta\, d\theta\, d\phi\, dt / \int dt$. The radial profiles are time-averaged over $68-73$ binary orbits. The mass accretion rate through the disk is highest at larger radii ($r \geq 300\,r_{\rm g}$). It dips to negative values between $200 - 300\,r_{\rm g}$, at the inner edge of the CBD. Here, the gas that does not acumulate in the overdensity is flung out by the binary torques, resulting in a negative total mass accretion rate. Below $200\,r_{\rm g}$ the remainder of the gas flows toward the binary with a net positive mass accretion rate, corresponding to 0.46\,$\dot{M}_{\rm Edd}$ in the MHD and 0.15\,$\dot{M}_{\rm Edd}$ in the RMHD run, reported earlier in this section.

\begin{figure}[t!]
\centering
  \includegraphics[trim=0 0 0 0, clip, width=1.0\linewidth]{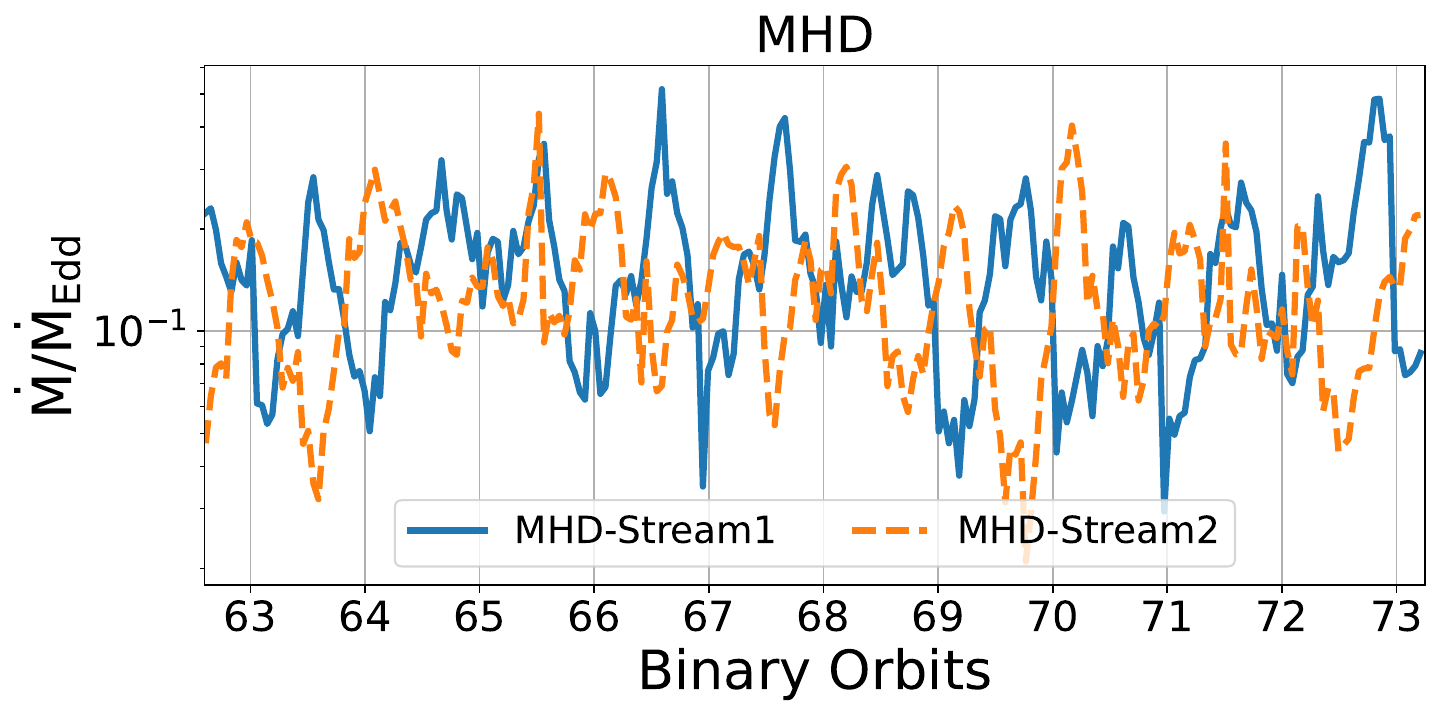}
  \includegraphics[trim= 0 0 0 0, clip, width=1.0\linewidth]{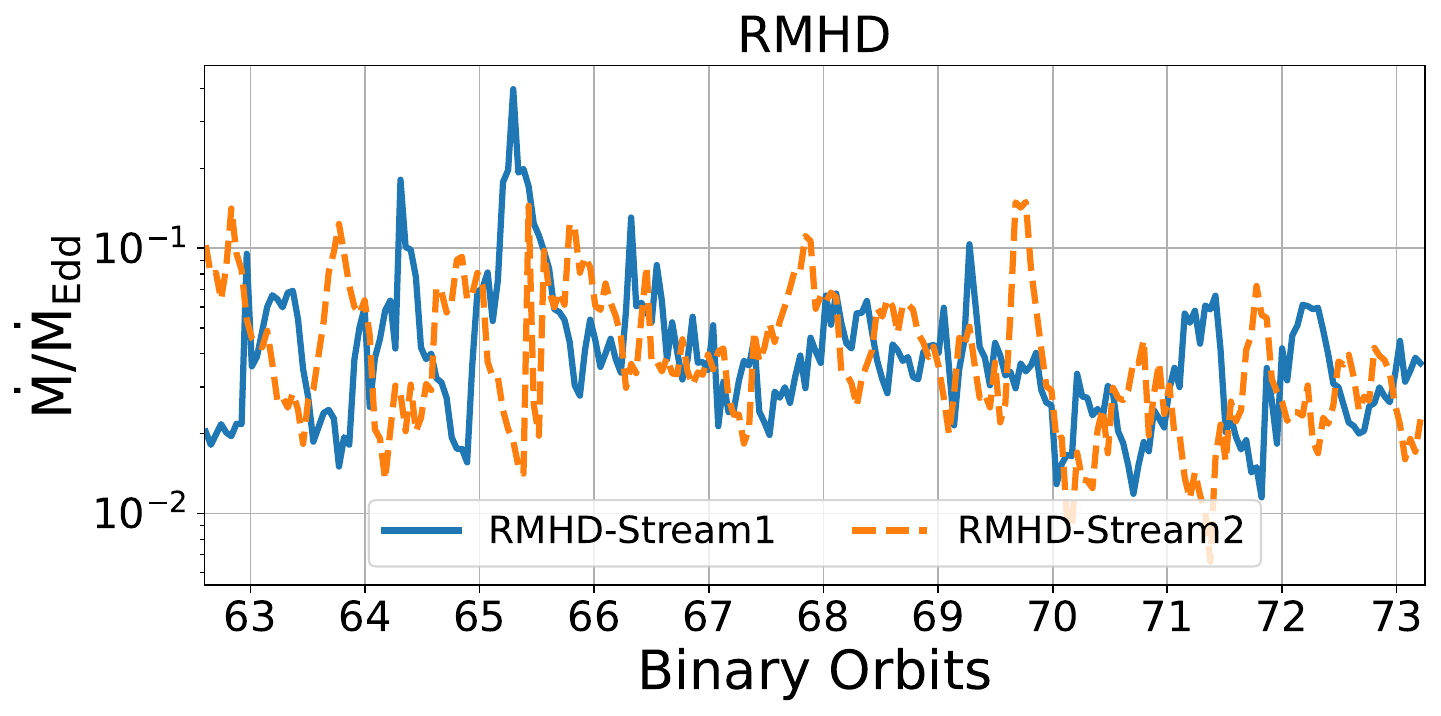}
\caption{Mass accretion rates through individual streams: stream1 (solid blue line) and stream2 (dashed orange) for both MHD (top) and RMHD (bottom) simulations for the last ten binary orbits. }
\label{fig:streams_Mdot}
\end{figure}

To determine how much of the accretion onto the binary is due to the streams, we use the definition of streams in Section~\ref{subsec:streams_cavity} and calculate the mass accretion rates via the stream falling on the first (stream-1) and the second (stream-2) MBH. The streams are equally strong at the start of the simulation. However, as the simulation progresses towards a quasi-steady state, the two streams begin to oscillate in strength, alternating every half a binary orbit. This oscillation is depicted in Figure~\ref{fig:streams_Mdot} and is consistent with findings by \citet{shi2012three}. The total mass accretion rate via the streams accounts for about 64\% of the mass accretion rate into the cavity for both MHD and RMHD simulations. This fraction is consistent with the contribution to the mass accretion rate from predominantly optically thick gas. Namely, Figure~\ref{fig:Mdot_Radial_MHD_vs_RMHD} shows that inside the disk cavity, contributions from both optically thick and optically thin gas phases are comparable. In the main body of the RMHD disk however, accretion rate is dominated by the optically thick phase, consistent with the single MBH accretion disk model AGN0.2 of \citet{jiang2019global}.

\section{Emission Properties}
\label{sec:obs_props}

\begin{figure*}[t]
\centering
  \includegraphics[trim=0 0 0 0, clip, width=0.49\linewidth]{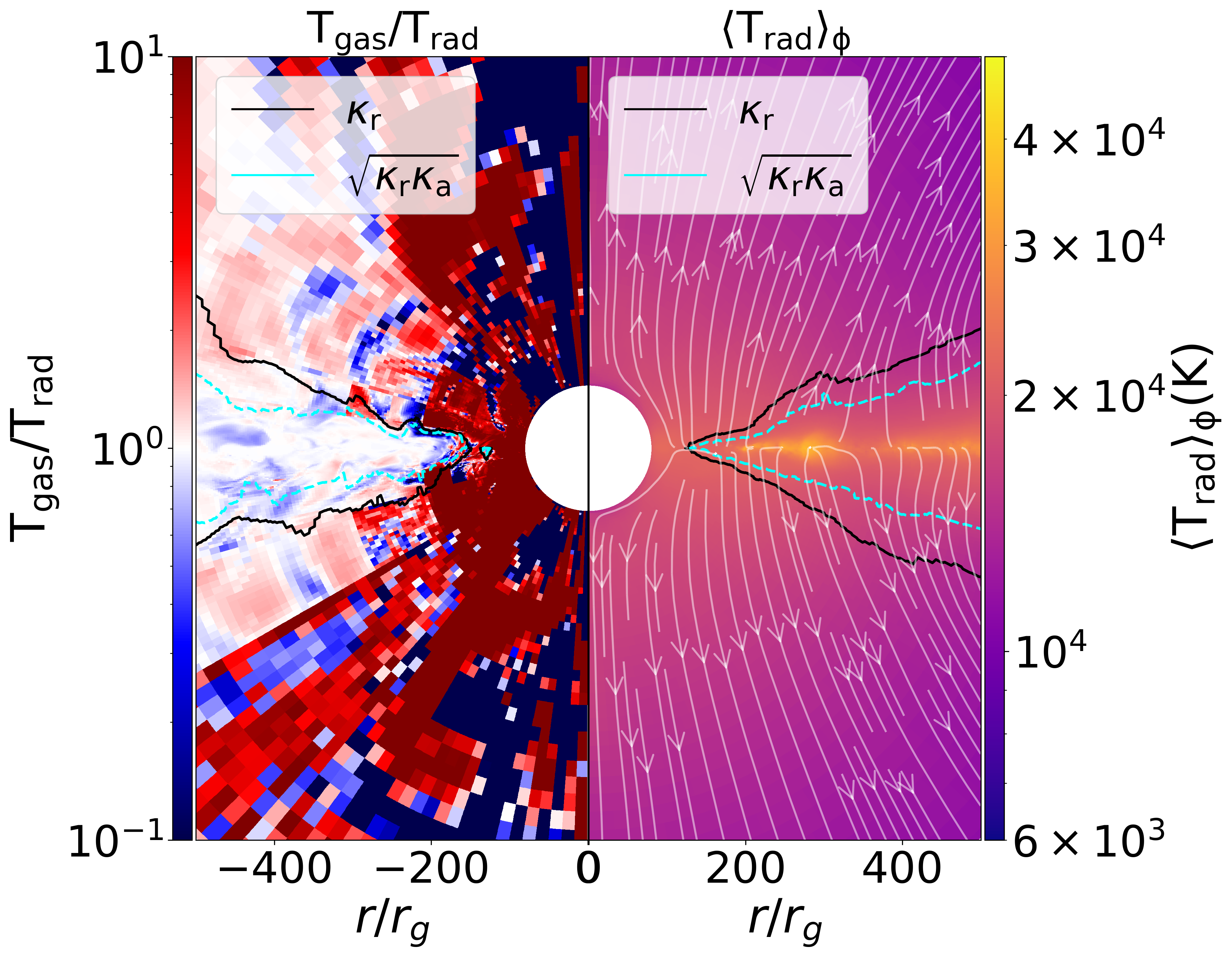}
 \includegraphics[trim=0 0 0 0, clip, width=0.49\linewidth]{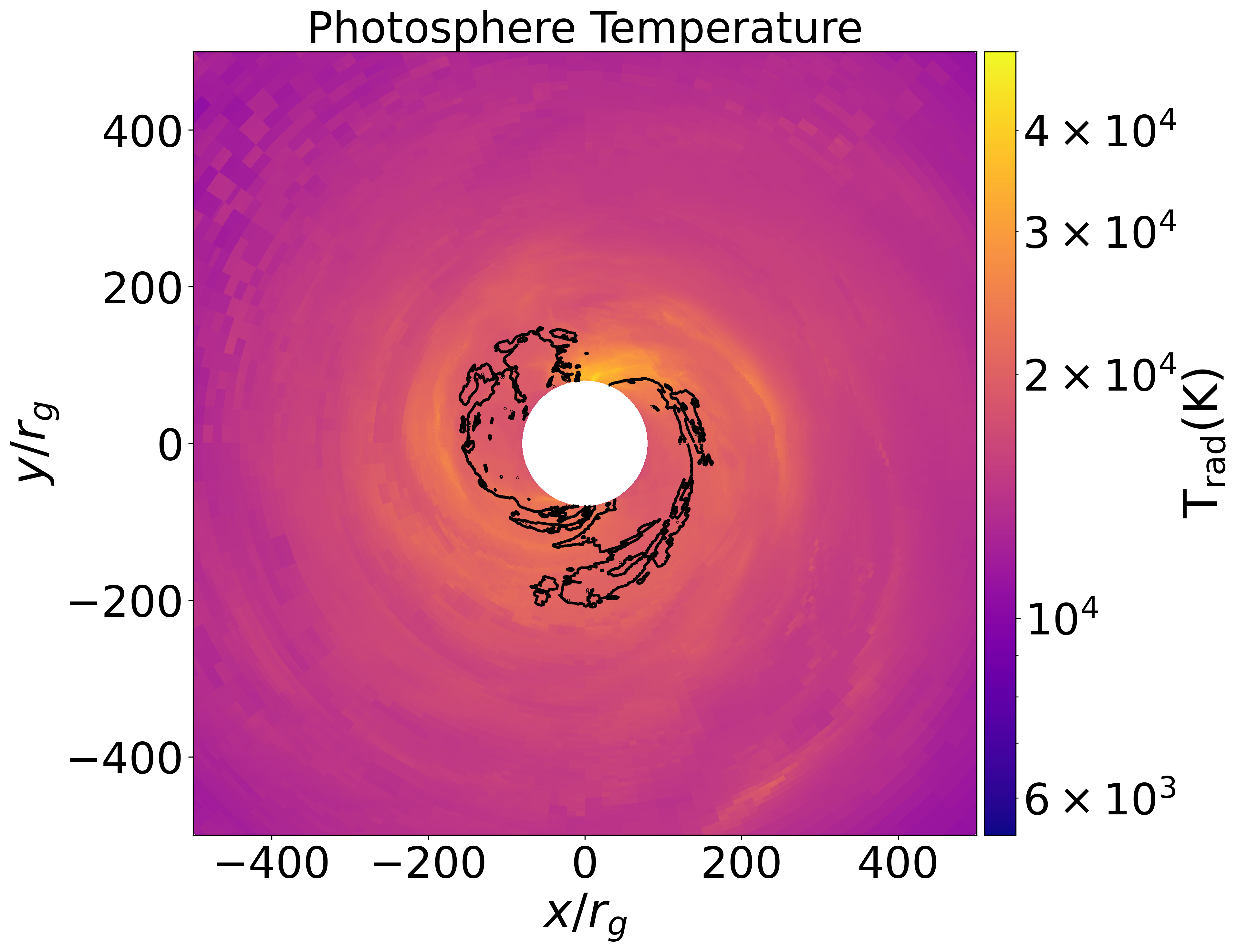}
\caption{{\it Left:} Slice of the ratio of gas to radiation temperature at $\phi = 0$ (left half-panel) along with azimuthally averaged radiation temperature (right half-panel) derived from the RMHD simulation of the disk. The lines show the location of the photosphere calculated using the Rosseland (black) and effective opacity (cyan). The streamlines illustrate the direction of radiation flux. {\it Right:} Radiation temperature calculated at the Rosseland photosphere. In the innermost regions where the photosphere is not defined (because $\tau < 1$), the figure shows the radiation temperature at the mid-plane of the disk.
}
\label{fig:PhotosphereDefn_Spectrum}
\end{figure*}

The radiative transfer module in Athena++ provides information about the frequency-averaged opacities of the gas, the radiation energy density, and radiation flux in the computational domain (see section~\ref{subsec:rmhd}). With this information, we identify the location of the photosphere, from which the emitted photons escape to ``infinity", as well as the properties of thermal emission of the gas from the surface of the circumbinary disk.

We calculate the thermal spectrum of the CBD using the radiation temperature at the photosphere, defined as $T_{\rm rad} = (E_{\rm r}\,c/4 \sigma)^{1/4}$, where $E_{\rm r}$ is the radiation energy density, and $\sigma$ is the Stefan-Boltzmann constant. To determine the location of the photosphere, we compute the optical depths from the poles ($\theta =0^{\circ}$ and $\theta = 180^{\circ}$) toward the disk mid-plane:
\begin{equation}
    \tau_i=\int \rho\, \kappa_i\, r\, d \theta \;,
\end{equation}
where $\kappa_i$ is either the Rosseland opacity, $\kappa_{\rm r}= \kappa_{\rm a} + \kappa_{\rm s}$, or the effective opacity, $\kappa_{\rm eff} = (\kappa_{\rm r} \kappa_{\rm a})^{1/2}$. Here, $\kappa_{\rm a}$ and $\kappa_{\rm s}$ are the Rosseland frequency-averaged absorption and scattering opacities, respectively. The photosphere is defined at the location where the optical depth $\tau_{\rm r}$ (using Rosseland opacity) or $\tau_{\text {eff }}$ (using effective opacity) equals unity. 

The left panel of Figure~\ref{fig:PhotosphereDefn_Spectrum} shows the locations of the Rosseland and effective photosphere, with the effective photosphere located slightly deeper in the disk atmosphere. Based on their definition, the Rosseland photosphere is a surface where photons last interact with plasma through absorption and scattering. The effective photosphere is defined as the last surface with $T_{\rm gas}/T_{\rm rad} = 1$, where photons thermally decouple from the gas, as shown in the figure. The left panel of Figure~\ref{fig:PhotosphereDefn_Spectrum} also depicts the radiation temperature and the direction of radiation flux with streamlines in the $r-\theta$ plane of the disk. The streamlines are calculated as $F_{r,\theta} = \langle F_{r} \rangle_{\phi} \,\hat{r} + \langle F_{\theta} \rangle_{\phi} \,\hat{\theta}$ and illustrate that photons escape radially out once they exit the photosphere. 

The right panel of Figure~\ref{fig:PhotosphereDefn_Spectrum} shows
that the radiation temperature at the Rosseland photosphere, in the upper hemisphere of the disk, is in the range $T_{\rm rad}\approx 2-3\times 10^4\,$K. The inner cavity is optically thin ($\tau < 1$), meaning that the photosphere is not defined in this region. We calculate the thermal spectrum using the Planck blackbody function
\begin{equation}
    L_{\nu} = \int \frac{2 \pi h \nu^3}{c^2}\frac{1}{e^{h \nu/k T_{{\rm rad}, i}} -1} \, dA \;,
\end{equation}
where $T_{{\rm rad},i}(r,\phi) $ is the radiation temperature at either the Rosseland or the effective photosphere. We do not take into account the emission from $\tau < 1$ regions.

\begin{figure}[t!]
\centering
 \includegraphics[trim= 0 0 0 0, clip, width=0.95\linewidth]{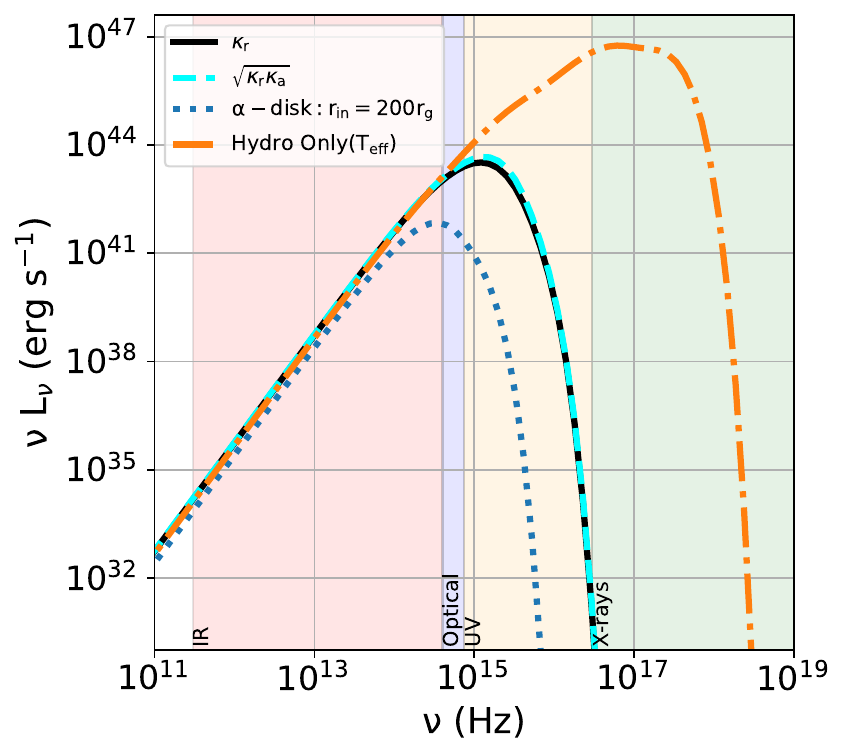}
\caption{Thermal spectrum calculated from the Rosseland (solid black) and effective opacity (dashed cyan), as described in Section~\ref{sec:obs_props}. For comparison, we also show the spectra calculated with two approximate methods commonly used in the literature: the $\alpha$-disk model (dotted blue) and the model using $T_{\rm eff}$ based on the hydrodynamic properties of the gas (dash-dot orange). All spectra were integrated to the radius set by the outer edge of the computational domain. See the text for more details.
}
\label{fig:Photosphere_EffectiveTemperature}
\end{figure}

Figure~\ref{fig:Photosphere_EffectiveTemperature} shows the spectra calculated from the Rosseland and effective photospheres. Since the effective photosphere lies somewhat deeper within the disk, where the radiation temperature is higher, its spectrum peak is slightly shifted toward higher frequencies. The difference between the two spectra is however negligibly small, indicating that $\kappa_a \gg \kappa_s$ and $\kappa_r \approx \kappa_a$. Hence, the thermal emission from the CBD, accreting at $ \sim 15 \%$ of the Eddington rate around an equal-mass MBHB of $2 \times 10^7 M_{\odot}$, peaks in the UV/optical band, with a luminosity $\nu L_\nu \approx 4.4\times10^{43}\,{\rm erg\,s^{-1}}$.

With the benefit of the thermal spectrum calculated from the first principles from our simulation with radiative transfer, it is worth examining how well it compares to the predictions of some of the approximate methods commonly used in the literature. For example, some of the earlier studies have computed thermal spectra by either assuming an $\alpha$-disk analytic model truncated at the cavity edge or by inferring the effective temperature from hydrodynamic properties of the gas in simulations. To compare our spectra with these methods, we calculate two additional spectra.

First, we compute the spectrum using the $\alpha$-disk model, whose temperature profile is given by Equation~\ref{eq:alpha_temp}:
\begin{equation}
\label{eq:alpha_temp}
    T(r)=\left\{\frac{3 G M \dot{M}}{8 \pi r^3 \sigma}\left[1-\left(\frac{r_{\rm in}}{r}\right)^{1 / 2}\right]\right\}^{1 / 4}
\end{equation}
where $r_{in}$ is the truncation radius, $M$ is the mass of the MBH and $\dot{M}$ is the mass accretion rate through the disk. For the purposes of this comparison, we adopt a point mass of $2 \times 10^7 M_{\odot}$ accreting at 15\% of the Eddington rate (assuming 10\% efficiency), and $r_{in} = 200 r_g$.

The second approximate spectrum, based on hydrodynamic properties of the gas in simulations, is calculated using the effective temperature defined by:
\begin{equation}
    T_{\rm eff}^4 = \frac{4}{3} \frac{\langle T_{\rm gas} \rangle_{\rho,\theta}^4}{\kappa_{\rm T}\,\Sigma}
\end{equation}
where $\langle T_{\rm gas} \rangle _{\rho,\theta}$ is the density-weighted vertical average of the gas temperature around the midplane and $\Sigma$ is the surface density of the disk. Note that in this approximation, the Thomson opacity $\kappa_{\rm T}\approx0.4\, {\rm cm^2\,g^{-1}}$ is assumed to be the dominant form of opacity, contrary to what we find from the radiative transfer calculation. This approach allows us to calculate the effective temperature using only the hydrodynamical quantities. As before, the regions with an optical depth of less than one are not used in the calculation.

The spectra derived from both approximate methods are shown in Figure~\ref{fig:Photosphere_EffectiveTemperature}.
The spectrum computed from the $\alpha$-disk model peaks in the optical and infrared band, at frequencies lower by a factor of a few compared to our baseline spectrum evaluated from the radiative transfer calculation. This difference arises because the $\alpha$-disk model assumes a single-MBH gravitational potential and doesn't capture the dynamics at the inner edge of the CBD, leading to a temperature profile that deviates substantially from the RMHD disk.

The approximate thermal spectrum computed from hydrodynamic properties of the gas and $T_{\rm eff}$ on the other hand peaks in the X-ray band, overestimating the peak frequency of thermal emission by two orders of magnitude. This discrepancy with our baseline spectrum arises because the gas temperature provides a poor approximation for the effective temperature of radiation at the inner edge of the CBD, where the gas and radiation are not in local thermodynamic equilibrium. In this region, the gas temperature is significantly higher than that of the radiation, resulting in the incorrect prediction of a much more energetic spectrum. These results underscore the importance of including radiation in simulations to obtain accurate emission properties that can be compared with observations.

\begin{figure}[t]
    \centering
    \includegraphics[width=0.45\textwidth]{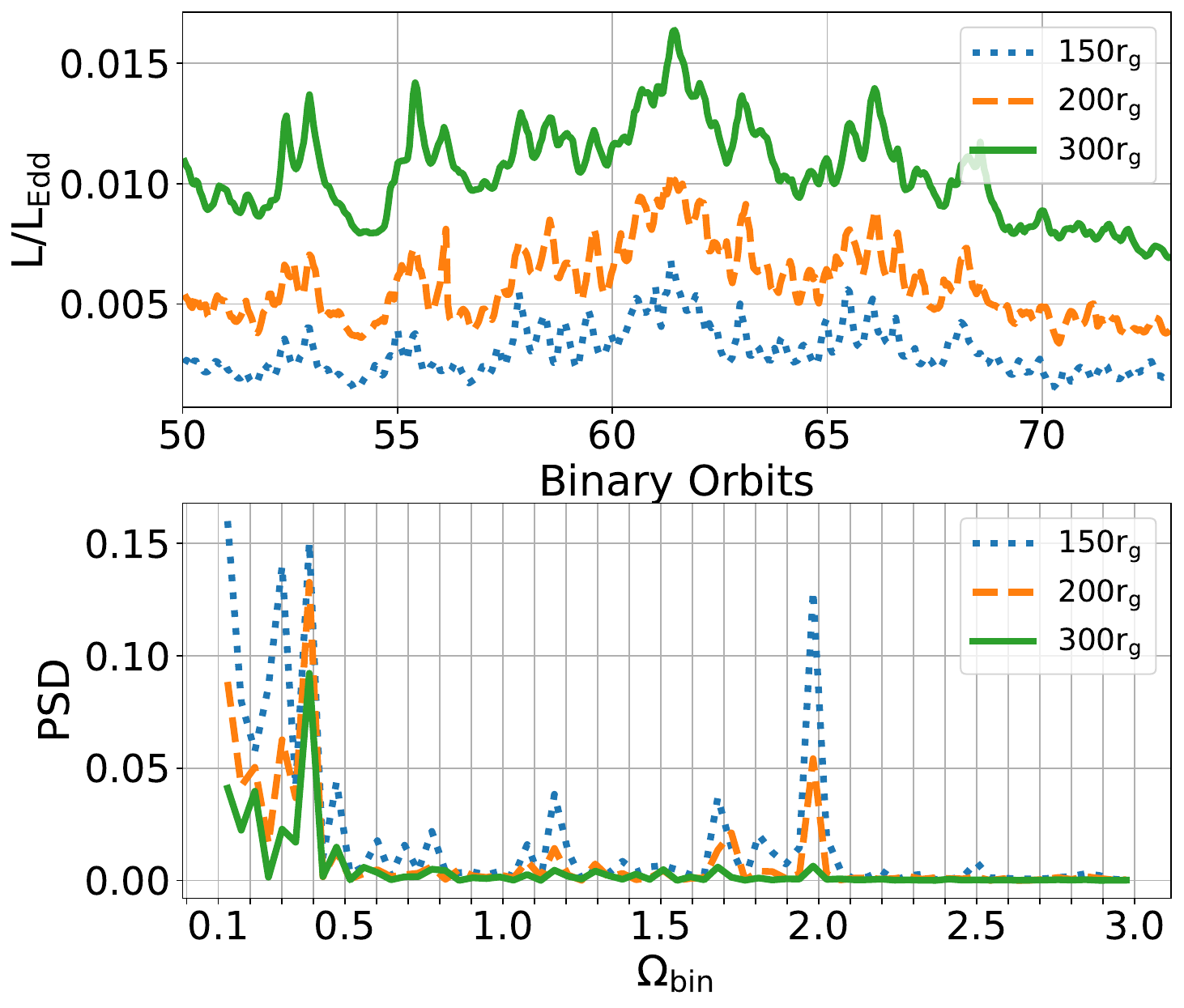}
    \caption{{\it Top:} Luminosity curves calculated at $r/r_{\rm g}$ = 150, 200, 300 from the RMHD run. {\it Bottom:} Power spectral density calculated after normalizing the light curves by their mean luminosity, showing two prominent modes: at 2\,$\Omega_{\rm bin}$ and at low-frequency 0.2-0.4\,$\Omega_{\rm bin}$. 
    }
    \label{fig:lightCurves}
\end{figure}

\begin{figure*}[t]
\centering
  \includegraphics[trim= 0 0 0 0, clip, width=0.45\linewidth]{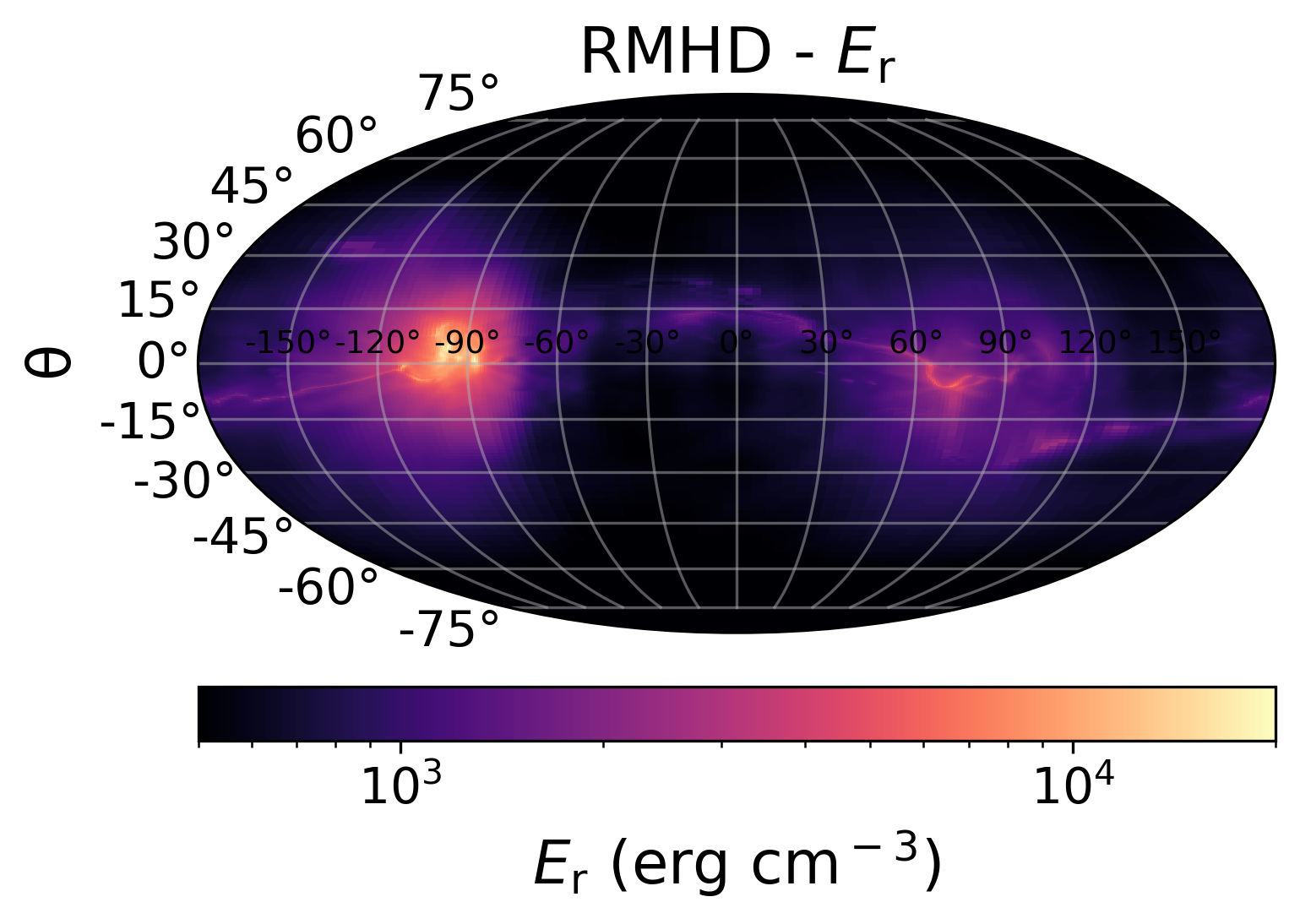}
     \includegraphics[width=0.48\textwidth]{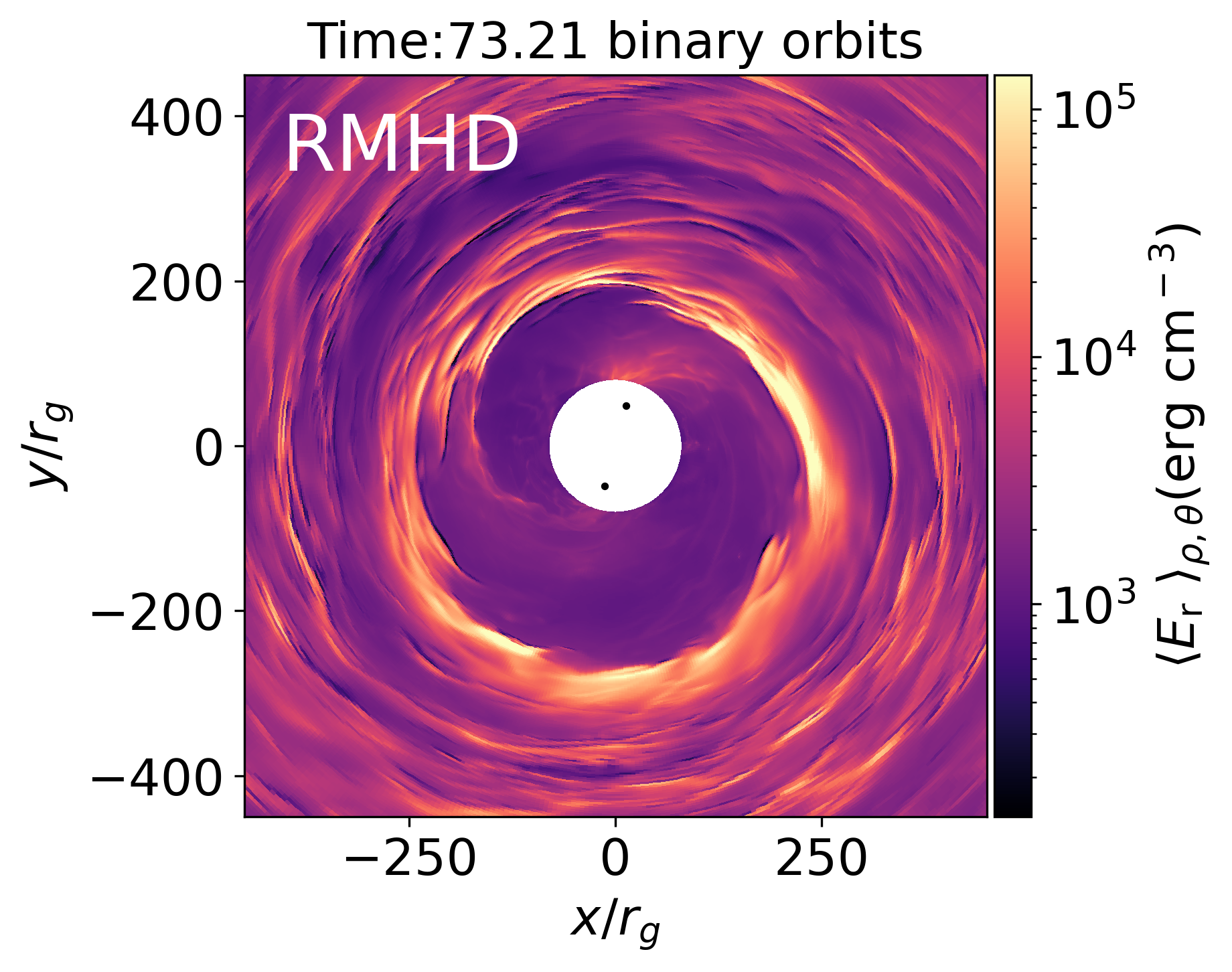}
\caption{{\it Left:} Spherical slice of radiation energy density at $r = 80\,r_g$ from the RMHD run at t$\approx$ 73 binary orbits. The stream with higher accretion rate (stream1) has higher radiation energy density than stream2. {\it Right:} Density weighted vertically averaged radiation energy density for the RMHD run at t $\approx$ 73 binary orbits. The streams flung-out by the binary shock and deposit energy at the inner edge of the CBD producing radiation.
}
\label{fig:streams_radiative}
\end{figure*}

Equipped with the thermal spectrum, in the next step we calculate the corresponding luminosity of radiation emitted by the circumbinary disk. We calculate the radiation luminosity from three concentric hemispheres (we choose $\theta < \pi/2$ hemisphere) located at the radii of $r/r_{\rm g} = 150, 200, 300$ by integrating the radial radiation flux over these hemispheres. This calculation includes the radiation flux from the optically thick and optically thin regions of the CBD that exits the hemisphere and excludes the inwardly advected radiation prevalent in the optically dense regions (see the streamlines in Figure~\ref{fig:PhotosphereDefn_Spectrum}).

The resulting light curves are presented in the top panel of Figure~\ref{fig:lightCurves}. The CBD contributes total luminosity of $\approx 0.01\,L_{\rm Edd}$ at $300\,r_{\rm g}$ and $\approx 0.018\,L_{\rm Edd}$ when calculated at the outer boundary of the domain, consistent with the peak luminosity inferred from the thermal spectrum. The power spectral density after normalizing the light curves by their mean luminosity is shown in the bottom panel of Figure~\ref{fig:lightCurves}. The variability of the luminosity depends on the radius and is influenced by the dynamics of the accretion streams and the inner edge of the CBD. At radii of 150\,$r_{\rm g}$ and 200\,$r_{\rm g}$, the luminosity exhibits variability associated with half the binary period, attributed to the emission from each stream becoming dominant once during each binary orbit (see section~\ref{subsec:mass_acc_rate}). This effect is illustrated in Figure~\ref{fig:streams_radiative}, showing the spherical slice of the radiation energy density at $r = 80\,r_{\rm g}$ from the RMHD simulation. Here stream1, which has a higher accretion rate, has a higher radiation energy density than stream2.

The emission from the overdensity at the inner edge also contributes to the modulation of luminosity, evident in the power spectral density distribution as a lower frequency mode $0.2-0.4\,\Omega_{\rm bin}$. We note that this lower-frequency mode is also present in the mass accretion rate that we calculate at the inner boundary ($r=80\,r_{\rm g}$), which is linked to the overdensity formed at the inner edge of the CBD, as discussed in section~\ref{subsec:mass_acc_rate}. At $r=300\,r_{\rm g}$ hemisphere, the variability associated with twice the binary orbital frequency diminishes and is diluted by contributions to the luminosity from the inner regions of the CBD. As a result, the modulation in the light curve caused by the energy dissipation at the inner edge of the CBD dominates. This effect can be seen in Figure~\ref{fig:streams_radiative}, which shows that the CBD's inner edge exhibits significantly higher radiation energy density than the rest of the disk because of the stream impacts.

\section{Discussion}
\label{sec:discussion}

Given that RMHD simulations are relatively computationally expensive and not yet a commonplace when it comes to simulations of accreting MBHBs (or even single MBHs), it is important to use them judiciously. In this section we discuss questions that particularly benefit from the RMHD approach (Section~\ref{subsec:implications}) and the aspects of circumbinary accretion where due to the limitations of our calculations other, established numerical methods are better suited (Section~\ref{subsec:limitations}).

\subsection{Implications for the Emission Properties of Circumbinary Disks}
\label{subsec:implications}

One advantage of the RMHD simulation presented here is that the location of the photosphere can be calculated directly, given the information about the density and opacity of the emitting gas. Combined with the information about the radiation temperature on this surface one can calculate thermal emission from the gas from first principles. These predictions provide an important benchmark for more approximate methods. As discussed in Section~\ref{sec:obs_props} and illustrated in Figure~\ref{fig:Photosphere_EffectiveTemperature}, the predicted thermal spectrum from the CBD can be off by orders of magnitude if the method does not capture the hydrodynamic effects, like shocks (in the case of analytic models), the opacity of the gas and the local properties of radiation (hydrodynamic models without radiative transfer). Therefore, the emission and radiative properties of the gas are the most important aspects that benefit from the calculation of radiative transfer. This includes not only the calculation of spectra but also the light curves and variability.

For example, thermal spectra calculated from our RMHD simulation (section \ref{sec:obs_props}) show that the thermal emission from the CBD peaks at around the optical/UV band for a $2\times10^{7}\,M_{\odot}$ separated by $100\,r_g$ and accreting at 0.15 times the Eddington rate. This is in contrast with studies that find soft X-ray emission at the inner edge from similar MBHB and disk configurations. The predicted UV emission can in principle be targeted by the next generation of UV observatories, like ULTRASAT \citep{shvartzvald2023ultrasat}, which will observe in the ultraviolet band (230–290 nm) and provide months-long light curves with minute-level cadence. Its sensitivity will allow it to target thermal emissions from the CBD around LISA binaries with mass $\sim 10^7\,M_{\odot}$, which will be bright enough to be detected to redshift $z\lesssim 0.7$, assuming a CBD luminosity of 0.01 $L_{\rm Edd}$.

The upcoming Vera Rubin observatory \citep{ivezic2019lsst} will be highly effective in detecting more massive sub-parsec binaries ($\sim10^8 - 10^9 M_{\odot}$), which are of interest to Pulsar Timing Arrays (PTAs). Its wide field of view, long observational baseline, and high cadence in the optical and near-infrared bands make it well-suited for targeting the thermal emission signatures from the circumbinary disks surrounding these binaries. For example, assuming that the effective temperature of the CBD scales with the binary mass as $\propto M_{\rm tot}^{-1/4}$, we estimate that the thermal emission from binaries with masses around $10^8 M_{\odot}$ and CBDs luminosity of $0.01\,L_{\rm Edd}$\footnote{Note however that the approximate extrapolation to higher MBHB masses does not take into account the differences in density and temperature of the CBDs that may arise from different opacity values, potentially altering the disk luminosity and spectrum.} will be detected in the optical bands up to a redshift of approximately 1.4. These systems would exhibit modulation in their light curve with periods of order 100 days, which are longer than the survey's cadence and fit well within the survey's 10-year projected observation period. More massive binaries, with masses around $10^9 M_{\odot}$, will also have their CBD thermal emission detectable in the optical and near-infrared filters of the Rubin Observatory. These systems would show much longer modulation periods, on the order of 1000 days, and can be observed up to a redshift of approximately 1.0. This makes the Vera Rubin observatory a powerful tool for studying massive black hole binaries and their circumbinary environments over significant cosmological distances. 

RMHD simulations determine the thickness of the CBD from first principles, unlike non-radiative hydrodynamic simulations that by necessity treat the disk's scale height as a free parameter. As discussed in Section~\ref{subsec:disk_structure}, for an MBHB with a total mass of $2 \times 10^{7}\,M_{\odot}$ accreting at 15\% of the Eddington rate, the scale height is calculated to be between 0.02 and 0.04, which is critical for the binary's orbital evolution. Previous HD studies by \citet{tiede2022binaries} and \citet{dittmann2022survey} have shown that CBDs with H/R less than 0.04 cause the binary orbit to shrink. Taking that outcome at the face value, the CBD disks around $\sim 10^7\,M_{\odot}$ equal-mass MBHBs accreting at 15\% of the Eddington rate will facilitate, rather than inhibit their mergers.

Furthermore, \cite{wang2023role} demonstrated that artificially increasing the cooling timescales in CBDs can cause their inner edge overdensity to disappear. They showed that transition from instantaneously cooled (isothermal) to more gradually cooled disks significantly affected the periodogram of the mass accretion rate. In our RMHD simulations, where emission of radiation self-consistently leads to cooling of the CBD, the overdensity persists, indicating that the cooling timescales are not long enough to eliminate it. As a result, the low frequency mode ($0.2-0.4\,\Omega_{\rm bin}$) is present in both the accretion rate and luminosity curve in our simulation.

\subsection{Limitations of the Calculation and Their Implications}
\label{subsec:limitations}

Our simulations span a finite length of about 75 binary orbits, which means that we can only reliably make predictions about physical phenomena that develop and evolve on characteristic timescales that are shorter than that. As noted before, 75 binary orbits correspond to about 9 orbits at a radius of $400\,r_{\rm g}$. We therefore expect that this simulation length is sufficient to reliably model physical properties and phenomena such as the disk density profile (including the overdensity), MRI activity, accretion rate, and emitted luminosity from the portion of the CBD within $400\,r_{\rm g}$. On the other hand, properties like precession of the inner edge of the CBD and torque calculations related to binary orbital evolution would require much longer simulation times, which would be computationally prohibitive. We therefore refrain from making predictions about these properties.

Our simulations do not account for general relativistic effects, which could in principle affect ($i$) the orbital dynamics of the fluid because of the differences in the shape of the assumed and real spacetimes, ($ii$) the evolution of the binary orbit due to the emission of GWs, and ($iii$) the propagation of emitted light rays. We expect these general relativistic effects to be minimal given that the separation between the MBHs and the bulk of the fluid (and its emitted radiation) is about $100\,r_{\rm g}$ or larger in our simulations, implying that the magnitude of general relativistic effects is $\lesssim 1\%$. Also, as noted in Section~\ref{subsec:physical_sys}, $t_{\rm orb}/t_{\rm gw} \lesssim 10^{-3}$ for the modeled MBHB configuration and we do not expect significant orbital evolution due to the emission of GWs over the span of 75 orbits.

We excise the central region within $80\,r_{\rm g}$, which includes the mini-disks around the individual MBHs and parts of the infalling streams. This excised region lies within the ``accretion horizon'', enclosed within the radius of approximately $100\,r_{\rm g}$, where hydrodynamic effects alone are unlikely to cause the gas to escape gravitational pull of the MBHs \citep{tiede2022binaries}. Thus, we expect our simulations to capture the salient properties of the flung-out streams, as they are launched by the MBHs from larger radii than that of the excised region. On the other hand, we do not capture radiation emitted from the mini-disks and the infaling portion of the streams. This radiation could influence the thermodynamic properties of the CBD by irradiating and heating its inner edge. A scenario in which this effect would be most pronounced is for unequal mass binaries with mass ratio $\approx 1/10$, where radiation emitted by the super-Eddington secondary mini-disk (associated with the less massive MBH) could drive substantial outflows. 

Our calculations of radiative transfer rely on approximate, frequency-averaged (gray) opacities. The impact that this may have on our simulations can be gleamed from the work by \citet{mills2024spectral},
who compared the outcomes of RMHD simulations of accretion disk around a single stellar-mass black hole
using gray opacity and frequency-dependent opacity. They found that RMHD simulations with grey opacity underestimate the gas temperature in the optically thin, funnel regions above and below the disk due to an inadequate treatment of Compton scattering.
This suggests that our RMHD simulation may not accurately capture Compton scattering in the optically thin regions as well. We plan to address this limitation in future work by calculating the multi-frequency radiative transfer in our simulations.

In this work, we focus on the thermal emission of the CBD and do not consider contributions to the luminosity from the other (nonthermal) emission mechanisms. Our predictions for luminosity and the spectrum are however subject to two important caveats. One is that the CBD emission, which would manifest itself as a part of the AGN emission, can be outshined by the host galaxy. This is particularly pertinent in lower-mass galaxies (such as those hosting LISA binaries) where AGN emission is regularly weaker than the host’s emission. The other caveat is that the UV emission from the CBD can be easily absorbed by the intervening matter in the host galaxy and the extragalactic space and may not be detectable as a result. Our simulations do not capture these effects.

\section{Conclusions}
\label{sec:con}

We present the first RMHD simulation of a sub-Eddington circumbinary disk around a $2 \times 10^7 M_{\odot}$ equal mass binary with orbital separation of $100\,r_{\rm g}$, that would be a precursor of a LISA GW source. We consider the thermodynamic properties and emission signatures of the CBD and compare them to our MHD simulation and similar works in the literature. Our most important findings are:

\begin{itemize}
    \item There are substantial differences in the structure of the CBD in the MHD and RMHD runs. In the former, the disk is predominantly supported by the thermal gas pressure, whereas in the latter it is supported by the magnetic pressure. The circumbinary disk in the RMHD run is a factor of $\sim 2$ thinner and denser and appears filamentary. 
    \item The accretion streams that form in the RMHD-simulated disk are weaker and more diffuse. They deliver a weaker impact at the inner rim of the CBD when flung out by the binary and as a result, the RMHD simulation exhibits a lower eccentricity CBD cavity and the overdensity that forms at the inner edge of the disk is less pronounced than the lump observed in our and other MHD runs.
    \item The accretion rate is $\sim 3$ times lower in the RMHD disk (corresponding to $0.15\,\dot{M}_{\rm Edd}$) because of the weaker stresses compared to the MHD run. In both runs streams supply about 60\% of mass accretion rate into the cavity. This fraction is consistent with contribution from predominantly optically thick gas. The rest of the accretion rate is contributed by the optically thin, diffuse gas filling the CBD cavity.
    \item The luminosity of the inner regions of the CBD is 1\% of $L_{\rm Edd}$, calculated as thermal radiation that peaks in the optical/UV band. The calculated light curve shows modulations corresponding to 0.5 and $3-5$ times the binary period (associated with the overdensity at the inner edge of the CBD). This indicates that despite the weaker overdensity and accretion streams in the RMHD run, the characteristic periodicities identified in the MHD runs persist.
    \item While the nature of radiative transfer calculations require us to pick a specific mass for the binary system, many of our findings are also applicable or can be easily generalized to the PTA binaries with mass in the range $10^8-10^{9}\,M_\odot$. The CBDs around such binaries will be optimal targets for the upcoming Vera Rubin observatory. 
\end{itemize}

In this work, we made a first step toward fully self-consistent simulations of accreting MBHBs and time-dependent predictions of their EM counterparts, based on radiative transfer calculations. These will be a stepping stone towards future, end-to-end simulations of MBHB inspiral and coalescence, which will link theory and observations in a direct and self-consistent way. In such simulations, the only free parameter necessary to describe a full set of EM signatures of an MBHB will be the accretion rate of the gas, in addition to the parameters needed for the description of their GWs. At that point, it will be possible to build a parametrized bank of the EM signatures, by exploring the parameter space, in a way similar to the GW templates. With enough targeted effort, substantial progress can be made in this area in the next few to ten years, just in time for the first detections of individual binaries by PTAs and before LISA is launched.

\begin{acknowledgments}
We thank the anonymous referee for helpful and constructive suggestions that improved our paper. This work was supported by the National Aeronautics and Space Administration (NASA) under grant 80NSSC19K0319, by the National Science Foundation (NSF) under grant AST-1908042, and by the Research Corporation for Science Advancement under award CS-SEED-2023-008. Research cyberinfrastructure resources and services supporting this work were provided in part by the NASA High-End Computing (HEC) Program through the NASA Advanced Supercomputing (NAS) Division at Ames Research Center, and in part by the Partnership for an Advanced Computing Environment (PACE) at the Georgia Institute of Technology, Atlanta, Georgia, United States of America.
\end{acknowledgments}

\vspace{5mm}


\software{Athena++ \citep{stone2020athena++,jiang2021implicit},  
          Matplotlib \citep{Hunter:2007}, 
          Numpy \citep{harris2020array},
          mpi4py \citep{dalcin2005mpi}
          }



\appendix

\section{Resolving the MRI}
\label{subsec:MRI_resolution}

\begin{figure}[ht!]
    \centering
    \includegraphics[width=1.0\textwidth]{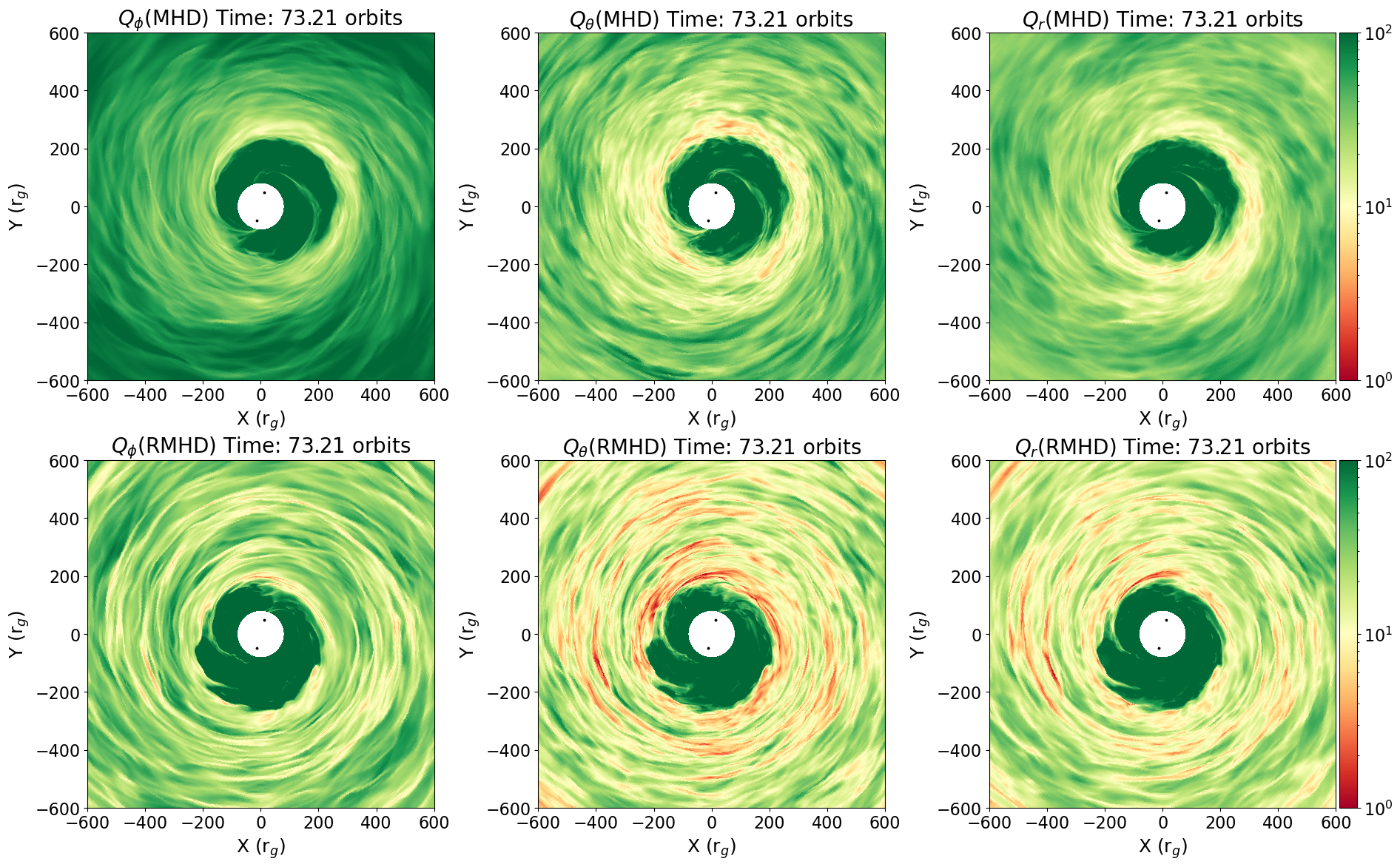}
    \caption{Density weighted vertically averaged quality factors measured at $t \approx 73$ binary orbits for the MHD (top) and RMHD (bottom) runs: $Q_{\phi}$ (left), $Q_{\theta}$ (middle), $Q_{r}$ (right).}
    \label{fig:vertical_avg_quality_factors}
\end{figure}

One of the most stringent criteria for numerical resolution in simulations involving magnetized accretion flows is set by the requirement to resolve the evolution of MRI, the instability responsible for driving accretion. To determine how well the MRI is resolved in our simulations, we calculate the characteristic MRI wavelength, $\lambda_{i} = (2 \pi \sqrt{16/15} |v_{A,i}| / \Omega)$, where $|v_{i}|$ is the Alfvén velocity along $i = \phi, \theta$ or $r$ and use it to evaluate the quality factors. The quality factors are the ratios of the grid cell size and the characteristic MRI wavelength $Q_{i} = \lambda_{i} / \Delta x_{i}$, where $\Delta x_{i}$ is $r \text{sin} \theta \Delta \phi$ along the azimuthal direction, $r \Delta \theta$ along the vertical, and $\Delta r$ along the radial direction. Figure \ref{fig:vertical_avg_quality_factors} shows the density-weighted vertical average for $Q_{\phi}, Q_{\theta}, Q_{r}$, for the MHD and RMHD runs.

We adopt the criteria for quality factors for our MHD and RMHD simulations commonly used for MHD simulations without radiation. From MHD studies, the properties of the MRI turbulence are expected to converge when $Q_{\phi}$ $\gtrsim$ 25 and $Q_{\theta}$ $\gtrsim$ 10 \citep{sorathia2012global,hawley2013testing}. In the main body of the MHD disk (away from the overdensity region), we resolve the quality factors in most regions at or above these threshold values. The quality factors are lower where gas density is higher and magnetic fields are weaker, such as in the lump in the MHD disk and in the filaments in the RMHD disk, similar to the findings reported by \citep{noble2021mass}. The lower quality factors evident at the inner edge of the CBD implies a slow growth of the MRI in that region. We expect this to have a minimal impact on the results of our simulations because at the inner edge, the gravitational torques from the binary dominate the angular momentum transport and are responsible for the mass accretion, rather than the stresses from the MRI turbulence.

We note that in the MHD run the mass-weighted $Q_{\phi}$ average in the region $400-500 r_g$ in the last 15 binary orbits (from $58-73$ binary orbits) decreases from 40 to 37 over the last 15 binary orbits (from $58-73$ binary orbits), while $Q_{\theta}$ decreases from 27 to 23. For the same run, average $Q_{\phi}$ decreases from 75 to 44.1 in the region $500-600 r_g$, while $Q_{\theta}$ evolves from 32 to 30. These quality factors suggest that we have enough cells to resolve MRI in the MHD run. For the RMHD run, the average $Q_{\phi}$ drops from 22 to 14 in the region $400-500 r_g$ and $Q_{\theta}$ from 14 to 8. In the outer region $500-600 r_g$ $Q_{\phi}$ drops from 34 to 21, while the average $Q_{\theta}$ drops from 20 to 13. These factors show that we have just about enough cells to resolve the MRI turbulence during the whole length of the RMHD simulations. The drop in the quality factors is however the most important criterion we used to decide when to terminate the RMHD simulation.

\bibliography{sample631}{}
\bibliographystyle{aasjournal}


\end{CJK*}
\end{document}